\documentclass[aps,prd,superscriptaddress,twocolumn,nofootinbib]{revtex4-2}
\usepackage{graphicx}
\usepackage[caption=false]{subfig}
\usepackage{epstopdf}
\usepackage{amsmath}
\usepackage{amsfonts}
\usepackage{amssymb}
\usepackage{latexsym}
\usepackage{hyperref}
\usepackage[english]{babel}
\usepackage[utf8]{inputenc}
\usepackage[colorinlistoftodos]{todonotes}
\usepackage{color}
\usepackage{slashed}
\usepackage{feynmp}
\usepackage{bm}
\usepackage{bbold}
\usepackage{eufrak}
\usepackage{slashed}
\usepackage{bm}  
\usepackage{xcolor}
\usepackage{footnote}
\usepackage{multirow}

\usepackage{ulem}

\begin{document}
	
	\title{Vortices in a parity-invariant Maxwell-Chern-Simons model}

	\author{W. B. De Lima}
	\email{wellissonblima@cbpf.br}
	\affiliation{Centro Brasileiro de Pesquisas F\'{i}sicas (CBPF), Rua Dr Xavier Sigaud 150, Urca, Rio de Janeiro, Brazil, CEP 22290-180}
	
	\author{P. De Fabritiis}
	\email{pdf321@cbpf.br}
	\affiliation{Centro Brasileiro de Pesquisas F\'{i}sicas (CBPF), Rua Dr Xavier Sigaud 150, Urca, Rio de Janeiro, Brazil, CEP 22290-180}

	\begin{abstract}
In this work we propose a parity-invariant Maxwell-Chern-Simons $U(1) \times U(1)$ model coupled with two charged scalar fields in $2+1$ dimensions, and show that it admits finite-energy topological vortices.
We describe the main features of the model and find explicit numerical solutions for the equations of motion, considering different sets of parameters and analyzing some interesting particular regimes. 	 
We remark that the structure of the theory follows naturally from the requirement of parity invariance, a symmetry that is rarely envisaged in the context of Chern-Simons theories.
Another distinctive aspect is that the vortices found here are characterized by two integer numbers.
	\end{abstract}
	

	\maketitle

	\pagestyle{myheadings}

	
	\section{Introduction} \label{sec_intro}
	
	Vortices are ubiquitous in nature, appearing from the rotating water in a sink to the winds surrounding a tornado. Such configurations can also be found throughout the physics literature, as illustrated in Refs~\cite{Vort1, Vort2, Vort3, Vort4, Vort5, Vort6, Babaev}.
	In field theory, vortices are defined as solitons and can appear whenever we have a continuous symmetry that is spontaneously broken and a vacuum manifold with a circular structure, as for example, in a (2+1)-dimensional abelian gauge theory in the Higgs phase~\cite{Shifman}.
	
	In this sense, the first appearance of vortices in the literature was in the context of superconductivity, through the work of Abrikosov in 1957~\cite{Abrikosov}. 
	In 1973, Nielsen and Olesen showed~\cite{NielsenOlesen} that the Abelian-Higgs (AH) model in 2+1 dimensions (the relativistic generalization of the Ginzburg-Landau model) admits finite-energy vortex solutions with a quantized magnetic flux. 
	An exact vortex solution was found by de Vega and Schaposnik in 1976~\cite{VegaSchaposnik}, considering the particular relation between the couplings for which scalar and vector bosons have the same mass.  
	The Abrikosov-Nielsen-Olesen (ANO) vortex described above is electrically neutral and, in fact, it was shown later by Julia and Zee in 1975~\cite{JuliaZee} that charged vortices with finite-energy cannot exist in the AH model. 
	
	A very interesting and subtle class of 2+1 topologically massive gauge theories was introduced in 1982 by Deser, Jackiw, and Templeton~\cite{DeserJackiwTempleton1, DeserJackiwTempleton2}, called nowadays Chern-Simons (CS) theories, after the pioneering work~\cite{ChernSimons} (see also Refs.~\cite{Schonfeld, JackiwTempleton, HagenAntigo1, HagenAntigo2}). 
	The CS term is exclusive of odd-dimensions, typically $\mathcal{P}$- and $\mathcal{T}$- odd, and topological in nature. In 2+1 dimensions, it gives a gauge invariant mass to the gauge field, providing a mass gap that cures the infrared divergences of these theories, changing drastically their physical content and leading to a quantization of the ratio between the CS parameter and the gauge coupling. 
	Over the years, CS theories have found applications all around physics, but the most famous breakthrough came with the work of Witten~\cite{Witten}, about the relationship between CS theories and the Jones polynomial. 
	For an introduction to CS physics, see Ref.~\cite{Dunne}; for a review of vortices in this context, see Ref.~\cite{Horvathy1}.
	
	It is well-known that a CS term has the property of flux attachment when coupled to matter fields, that is, it relates the electric charge with the magnetic flux.
	In 1986, it was shown that finite-energy charged vortices solutions exist in Abelian~\cite{PaulKhare} and non-abelian~\cite{KumarKhare, VegaShap1, VegaShap2} Higgs models in the presence of a CS term (see also Ref.~\cite{Pisarski}); the existence of quantum charged vortices has been shown in Ref.~\cite{Frohlich}. Interestingly enough, charged vortices can play an important role in condensed matter, for example, in the fractional quantum Hall effect~\cite{Laughlin}, high-$T_c$ superconductors~\cite{ChenWilczek}, and superfluids~\cite{VolovikYakovenko}.
	
	In the pure CS limit, when the Maxwell kinetic term is absent, peculiar charged vortices were shown to exist~\cite{JatkarKhare}, with magnetic field vanishing at the origin, instead of taking a finite value as usual.
	An interesting work studying vortices in a Maxwell-Chern-Simons-Higgs model, interpolating between AH model and pure CS-Higgs case was done in Ref.~\cite{Boyanovsky}.  
	Upon choosing a suitable potential, it was shown in Refs.~\cite{HongKimPac, JackiwWeinberg} that it is possible to obtain a Bogomol'nyi-type~\cite{Bogomol'nyi} energy lower bound with first order equations that describe self-dual topological charged vortices in the Higgs phase of the Chern-Simons-Higgs model. 
	We remark that there are non-topological solitons with non-zero flux in the symmetric vacuum~\cite{JackiwLeeWeinberg}. 
	Since Supersymmetry and self-duality are intimately related~\cite{WittenOlive,DiVecchia,Hlousek1,Hlousek2}, a $\mathcal{N}=2$ supersymmetric extension is possible~\cite{LeeLeeWeinberg} (see also~\cite{LeeLeeMin2,Edelstein,Alvaro}).
	In Ref.~\cite{LeeLeeMin} the authors studied topological and non-topological vortices in self-dual models with both Maxwell and Chern-Simons terms; for more details on self-dual CS theories, one can see Ref.~\cite{Dunne2}. 
	This kind of soliton solutions can also be found in non-relativistic theories (see, for instance, Refs.~\cite{JackiwPi1, JackiwPi2, Manton, Horvathy2}).
	
	It is usually said that the presence of a CS term necessarily causes the violation of $\mathcal{P}$ and $\mathcal{T}$ symmetries. Although usually correct, this is not always true. In fact, it was already pointed out in \cite{DeserJackiwTempleton1, DeserJackiwTempleton2} and later shown by Hagen~\cite{Hagen}(see also Ref.~\cite{Wilczek}), that a gauge and parity-invariant CS theory can be constructed by essentially doubling the gauge degrees of freedom and adopting their respective CS terms with opposite signs. 
	A different approach was proposed by Del Cima and Miranda~\cite{Oswaldo1} a few years ago in the context of graphene physics (see also Ref.~\cite{Gorbar}). The authors introduced a parity-preserving $U(1) \times U(1)$ massive quantum electrodynamics (QED) with two gauge fields having different behaviors under parity, and a CS term mixing them, a distinctive feature of the model. Its massless version was studied in Ref.~\cite{Wellisson1}, and it was shown that it exhibits quantum parity conservation at all orders in perturbation theory~\cite{Oswaldo2}. Recently, it was shown in Ref.~\cite{Wellisson2}, that the massive version is ultraviolet finite, that is, exhibits vanishing $\beta$-functions associated to the gauge coupling constants and CS parameter, and also vanishing anomalous dimensions. Furthermore, it was shown that the model is parity and gauge anomaly free at all orders in perturbation theory.

	Vortices in this context have already been discussed in the literature. In Ref.~\cite{Kim}, the authors studied vortices in a $U(1) \times U(1)$ CS model coupled with scalar matter exhibiting fractional and mutual statistics. Following this work, the low energy dynamics of vortices was investigated in~\cite{Dziarmaga1} (see also~\cite{Shin2}), hybrid anyons in~\cite{Dziarmaga2}, and vortices in a CS theory coupled with fermions in~\cite{Shin1}. These works had as a background experiments sugesting parity-invariance in high-$T_c$ superconductors~\cite{Exp1, Exp2,Exp3}, and the subsequent theoretical models agreeing with them~\cite{Semenoff,Mavromatos,Kovner,Mavromatos2}. Finally, this subject is also investigated in the mathematical physics literature~\cite{Math1, Math3, Math4}, and interestingly enough, similar models with a mixed CS term find many applications in condensed matter~\cite{Kou1, Kou2, Kou3, Kou4, Qi, Ye, Diamantini1, Diamantini2, Diamantini3, Sakhi}.

	In the last few years, there have been several contributions to the literature of vortices, and here we briefly mention some of them. In Ref.~\cite{Shaposhnikov}, the authors reported a new topological vortex solution in a $U(1) \times U(1)$ Maxwell-Chern-Simons theory. Considering the situation in which one of the $U(1)$'s was spontaneously broken, they obtained a long-range force, protected at the quantum level by the Coleman-Hill theorem~\cite{ColemanHill}. Another interesting development was achieved in Refs.~\cite{Penin1, Penin2}, where the authors used a systematic expansion in inverse powers of $n$ to study giant vortices with large topological charge, observed experimentally in condensed matter systems~\cite{GiantExp1, GiantExp2, GiantExp3}.
	In Ref.~\cite{Helayel}, the authors considered a $U(1) \times U(1)$, $\mathcal{N} = 2$ supersymmetric model in $2+1 $ dimensions, investigating magnetic vortex formation and discussing applications of it. For some recent developments on vortex solutions within the gravitational context, see for instance Refs.~\cite{Edery, Albert}.
	Other interesting recent works can be found in Refs.~\cite{Schaposnik1, Schaposnik2, Schaposnik3, Bazeia1, Bazeia2, Bazeia3}.
	
	In this work we propose a parity-invariant Maxwell-Chern-Simons $U(1) \times U(1)$ scalar QED in 2+1 dimensions, in analogy with the fermionic matter case studied in Ref.~\cite{Oswaldo1}, and investigate the existence of topological vortices in the Higgs phase of this model. 
	Although vortices in similar scenarios have already been considered in the literature, they have been restricted to the pure CS case. The addition of a Maxwell term, more than an academic exercise, leads to physically sensible differences, changing for example the number of propagating degrees of freedom, the quantization procedure and even the nature of the vortices themselves. Moreover, the pure CS limit can in principle be achieved by a suitable choice of parameters in a Maxwell-CS model,
	\footnote{In Ref.~\cite{Nemeth} the author advocates that the pure CS limit does not describe the large distance limit of the Maxwell-CS model.} but the converse is certainly not true. Therefore, this work comes as one more step towards the description of physical phenomena where charged vortices or anyonic matter may play an important role while preserving $\mathcal{P}$ and $\mathcal{T}$.

	This paper is organized as follows: In Sec.~\ref{sec_theory}, we present the model and build the theoretical setup, introducing its field content, symmetries, and also the scalar potential we shall be working with. We show how charge and flux are related in our parity-invariant theory and present the mass-spectrum around the symmetric vacuum of the potential. In Sec.~\ref{sec_topology} we discuss general properties of the topological configurations considered here such as charge, flux and angular momentum quantization due to the boundary conditions, we comment on the expected asymptotic behavior of the solutions and we cast the static equations of motion on a more suitable form for numerical investigations. We present explicit vortex solutions in Sec.~\ref{sec_vortex} and discuss the main features of the scalar profiles, electric, g-electric, magnetic, and g-magnetic fields, followed by an analysis of their dependence on the parameters of the theory. We also evaluate some physical quantities associated with each solution such as charge, g-charge, fluxes and angular momentum. The static energy (mass) of solutions and those with half-integer fluxes are briefly commented. The analysis of limiting cases (pure Maxwell and pure CS) is done in Sec.~\ref{AppendixB}. Finally, in Sec.~\ref{sec_conclusions}, we state our concluding remarks. We use natural units ($c = \hbar = 1$) and the flat Minkowski metric $\eta^{\mu\nu}={\rm diag}(+1,-1,-1)$, throughout; for the Levi-Civita tensor, we use the conventions: $\epsilon^{0 1 2} = -1$, $\epsilon^{i j} \equiv \epsilon^{0 i j}$, where Latin indices always refer to spatial components.

	\section{Theoretical Setup} \label{sec_theory}
	
	Let us propose a parity-invariant Maxwell-Chern-Simons $U(1)_A \times U(1)_a$ scalar QED in 2+1 dimensions with Lagrangian given by
	\begin{align}\label{LagInicial}
		\mathcal{L} = &-\frac{1}{4} F_{\mu \nu} F^{\mu \nu} -\frac{1}{4} f_{\mu \nu} f^{\mu \nu}  + \mu \epsilon^{\mu \nu \rho} A_\mu \partial_\nu a_\rho  \nonumber \\
		&+ \vert D_\mu \phi_+ \vert^2 + \vert D_\mu \phi_- \vert^2 - V\left(\vert \phi_+ \vert, \vert \phi_- \vert\right),
	\end{align}
	where the covariant derivative with respect to the gauge group $U(1)_A \times U(1)_a$ acting on the complex scalar fields $\phi_+$ and $\phi_-$ is given by
	\begin{align}
		D_\mu \phi_\pm = \partial_\mu \phi_\pm +i e A_\mu \phi_\pm \pm i g a_\mu \phi_\pm.
	\end{align}
	In the above expression, $e$ and $g$ are the gauge couplings associated with the gauge groups $U(1)_A$ and $U(1)_a$, respectively, and $\mu > 0$ is the CS parameter. The field strength tensors are given by $F_{\mu \nu} = \partial_\mu A_\nu - \partial_\nu A_\mu$ and $f_{\mu \nu} = \partial_\mu a_\nu - \partial_\nu a_\mu $, respectively. Notice that the scalar fields have the same charge under $U(1)_A$ but opposite charges under $U(1)_a$. The mass dimensions here are: $\left[e^2\right] =\left[g^2\right] = \left[\mu\right] = 1$ and $\left[A_\mu\right] = \left[a_\mu\right] = \left[\phi_\pm\right] = 1/2$. In this model, in analogy with the fermionic version studied in Ref.~\cite{Oswaldo1}, the gauge field $a_\mu$ is a pseudo-vector under parity, and its presence in the mixed CS term is what allows a CS theory to be parity-invariant. 
	
	The Lagrangian presented here is by construction invariant under $U(1)_A \times U(1)_a$ gauge transformations:
	\begin{align}\label{GaugeTrans}
		\phi_\pm'(x) &= e^{i \left( \rho(x) \pm \xi(x)\right)} \phi_\pm(x), \nonumber \\
		A_\mu'(x) &= A_\mu(x) - \frac{1}{e} \partial_\mu \rho(x), \nonumber \\
		a_\mu'(x) &= a_\mu(x) - \frac{1}{g} \partial_\mu \xi(x).
	\end{align}
	
	To ensure parity-invariance of this model, the scalar fields should behave somehow in the same way under parity as the fermionic matter in Ref.~\cite{Oswaldo1}. Thus, we will extend the parity concept to include a transformation in the space of fields that swaps the role of $\phi_\pm$:
	\begin{align}
		A_\mu^P &= \mathcal{P}_\mu^{\; \nu } \, A_\nu, \nonumber \\
		a_\mu^P &= - \mathcal{P}_\mu^{\; \nu} \, a_\nu, \nonumber \\
		\phi_\pm^P &= \eta \, \phi_\mp, 
	\end{align}
	where we have $\mathcal{P}_\mu^{\; \nu} = diag(+-+)$, and $\eta$ is a complex phase. 
	One can immediately see that, with these transformations, and assuming that a suitable potential $V$ is chosen, our model is parity-invariant. 

	The most general renormalizable potential compatible with the symmetries of the model is
	\begin{align}\label{key}
		V &= m^2 \left(\vert \phi_+\vert ^2 + \vert \phi_-\vert ^2  \right) + \frac{M_1}{2} \left(\vert \phi_+\vert ^4 + \vert \phi_-\vert ^4 \right) \nonumber \\  
		& + M_2 \vert \phi_+\vert ^2  \vert \phi_-\vert ^2  + \frac{g_1}{3}\left(\vert \phi_+\vert ^6 + \vert \phi_-\vert ^6  \right)
		\nonumber	\\  
		& + g_2 \left(\vert \phi_+\vert ^2 \vert \phi_-\vert^4 + \vert \phi_-\vert ^2 \vert \phi_+\vert^4 \right),
	\end{align}
	where the parameters should be carefully chosen in order to ensure the presence of only stable vacua. It should be clear that, depending on the parameters, different vacua structures might appear, which could in principle lead to the spontaneous breaking of one, both, or none of the $U(1)$ symmetries.
	Let us choose the simplest scalar potential  that leads to a spontaneously broken but parity-symmetric vacuum. Thus, we will consider, with $\lambda >0$:
	\begin{align} 
		V\left(\phi_+, \phi_-\right) = \frac{\lambda}{4} \left( \vert \phi_+ \vert^2 - v^2 \right)^2 + \frac{\lambda}{4} \left( \vert \phi_- \vert^2 - v^2 \right)^2.
	\end{align}
	
	This is the simplest extension of the Abelian-Higgs potential for the case under study. Taking $v \neq 0$, it will clearly induce a non-trivial vacuum expectation value (VEV) for the scalar fields, putting the theory into the Higgs phase, where we have $\langle |\phi_\pm | \rangle = v$. This potential is not stable under quantum corrections, but this will not be an issue, since we are focusing on classical solutions.
	
	An important remark must be made at this point. If one defines the fields $A_\mu^\pm = \left(A_\mu \pm a_\mu \right) / \sqrt{2}$, the pure gauge part of the Lagrangian would be rewritten as
	\begin{align} 
		\mathcal{L} \supset &-\frac{1}{4} F_{\mu \nu}^+ F^{\mu \nu +} -\frac{1}{4} F_{\mu \nu}^- F^{\mu \nu -} \nonumber \\
		&+ \frac{\mu}{2} \epsilon^{\mu \nu \rho} \left( A_\mu^+ \partial_\nu A_\rho^+ - A_\mu^- \partial_\nu A_\rho^- \right),  
	\end{align}
	realizing the parity-invariance of the model in a different, although equivalent, form, as studied in Ref.~\cite{Hagen}. The other part of the Lagrangian in this setting is written as
	\begin{align} 
		\mathcal{L} \supset &\vert \left(\partial_\mu + i q_1 A_\mu^+ + i q_2 A_\mu^-\right) \phi_+ \vert^2 \nonumber \\
		+ &\vert \left(\partial_\mu + i q_2 A_\mu^+ + i q_1 A_\mu^-\right) \phi_- \vert^2 - V\left(\vert \phi_+ \vert, \vert \phi_- \vert \right).
	\end{align}
	
	In the above expression, one can see that $\phi_+$ and $\phi_-$ have swapped effective charges, defined as $q_1 = \left(e + g\right)/\sqrt{2}$ and $q_2 = \left(e - g\right)/\sqrt{2} $. The parity transformation is realized by $A_\mu^{\pm} \rightarrow \mathcal{P}_\mu^{\,\,\nu} A_\nu^{\mp}$ and $\phi_\pm \rightarrow \eta \phi_\mp$. This setup explicitly exhibits the parity-invariance of the CS sector. It is possible to show that, the on-shell free fields $A_\mu^+$ and $A_\mu^-$ provide the vector representations of the three-dimensional Poincaré group with spins equal to +1 and -1 (for $\mu > 0$), respectively, as one can see in Ref.~\cite{Binegar}. In this paper, although, we have chosen to work with the variables  $A_\mu$ and $a_\mu$ for convenience.

	The equations of motion following from the Lagrangian are given by
	\begin{align}\label{EOM}
		\partial_\mu F^{\mu \nu} + \mu \epsilon^{\nu \alpha \beta} \partial_\alpha a_\beta &= e \left(J_+^\nu + J_-^\nu\right), \nonumber \\
		\partial_\mu f^{\mu \nu} + \mu \epsilon^{\nu \alpha \beta} \partial_\alpha A_\beta &= g \left(J_+^\nu - J_-^\nu\right), \nonumber \\		
		D_\mu D^\mu \phi_\pm &= -\frac{dV}{d\phi_\pm^*},
	\end{align}
	where the currents are $J_\pm^\nu = i \left[ \phi_\pm^* D^\nu \phi_\pm - \phi_\pm D^\nu \phi_\pm^*  \right]$.
	
	Let us take a look at the peculiar Gauss laws that this model presents. Define the electric and magnetic fields associated with the gauge fields $A_\mu $ and $a_\mu$ by $E^{i} = F^{i 0}$, $B = \epsilon^{i j} \partial_i A_j$ and $e^{i} = f^{i 0}$, $b = \epsilon^{i j} \partial_i a_j$, respectively. From the gauge fields equations of motion, and using $\rho_\pm = J^0_\pm$: 
	\begin{align}\label{GaussLaw}
		\vec{\nabla} \cdot \vec{E} + \mu b = e \left(\rho_+ + \rho_- \right), \nonumber \\
		\vec{\nabla} \cdot \vec{e} + \mu B = g \left(\rho_+ - \rho_- \right).
	\end{align}
	
	Defining the electric charge $Q = e \int \! d^2x \, \left(\rho_+ + \rho_-\right)$ and the g-electric charge $G = g \int \! d^2x \, \left(\rho_+ - \rho_-\right)$, and defining also the magnetic flux as $\Phi \equiv  \int \! d^2x \,  B $ and the g-magnetic flux as $\chi \equiv  \int \! d^2x \,  b$, we obtain upon integration:
	\begin{align} 
		Q = \mu \chi, \quad 
		G = \mu \Phi.
	\end{align}
	That is, the electric charge associated with one gauge field is proportional to the magnetic flux associated with the other. It is well-known that there is a flux attachment caused by the CS term, but in our case this charge-flux relation happens between two different gauge fields. This mutual statistics behavior~\cite{Wilczek} is a distinctive feature of this class of models~\cite{Kim}, but here we implement the flux attachment in a parity-invariant way.
	
	The energy-momentum tensor here can be written as
	\begin{align}\label{emtensor}
		T^{\mu\nu}
		&=  \left(\eta^{\mu\nu}\frac{1}{4} F_{\alpha \beta}F^{\alpha \beta}  -  F^{\mu\beta}F^{\nu }_{~~\beta}\right) \nonumber 
		\\ 
		&+  \left(\eta^{\mu\nu}\frac{1}{4} f_{\alpha \beta}f^{\alpha \beta}  -  f^{\mu\beta}f^{\nu }_{~~\beta}\right)
		\nonumber \\   &+D^\mu \phi^*_+ D^\nu \phi_+ + D^\mu\phi_+ D^\nu \phi^*_+   -\eta^{\mu\nu}\vert D_\alpha \phi_+ \vert^2  \nonumber \\ &+D^\mu \phi^*_- D^\nu \phi_- + D^\mu\phi_- D^\nu \phi^*_- -\eta^{\mu\nu} \vert D_\alpha \phi_- \vert^2 \nonumber\\
		&+ \eta^{\mu\nu} V.
	\end{align}
	The energy functional following from this expression is
	\begin{align}\label{Energy}
		E = &\int \! d^2x \left[ \frac{1}{2}\left(\vec{E}^2 + B^2\right) + \frac{1}{2}\left(\vec{e}^2 + b^2\right) + V \right. \nonumber \\     
		&\left. + \vert D_0 \phi_+ \vert^2 + \vert D_0 \phi_- \vert^2 + \vert D_i \phi_+ \vert^2 + \vert D_i \phi_- \vert^2  \right].
	\end{align}
	We are interested only in the static regime, {\it i.e.}, $\partial_0 \equiv 0$.

	The vacuum configuration of the system is given by the absolute minimum of the energy functional, that can be achieved, for instance, considering $\phi_\pm  = v$ and $  A_\mu  =  a_\mu  = 0$. In the unitary gauge we can write $\phi_\pm(x) = v + h_\pm(x)/\sqrt{2}$. The quadratic part of the Lagrangian here is given by
	\begin{align} 
		\mathcal{L}^{\text{quad}} &= -\frac{1}{4} F_{\mu \nu} F^{\mu \nu} - \frac{1}{4} f_{\mu \nu} f^{\mu \nu} + \frac{1}{2} (\partial_\mu h_+)^2 +  \frac{1}{2} (\partial_\mu h_-)^2 \nonumber \\
		&+  2v^2 \left(e^2 A_\mu A_\mu + g^2 a_\mu a_\mu \right) - \frac{\lambda v^2}{2} (h_+^2 + h_-^2) \nonumber \\
		&+ \frac{\mu}{2} \epsilon^{\mu \nu \rho} A_\mu \partial_\nu a_\rho +  \frac{\mu}{2} \epsilon^{\mu \nu \rho} a_\mu \partial_\nu A_\rho.
	\end{align}
	From the above expression we can immediately see that we have two degenerate massive scalars with $m_S = \sqrt{ \lambda v^2}$. For the gauge quadratic part we can write
	\begin{align} 
		\mathcal{L}_{\text{gauge}}^{\text{quad}} = 
		\frac{1}{2} \left( \begin{array}{cc}  A_\mu \quad a_\mu \end{array}\right) O^{\mu \nu} \left( 
		\begin{array}{c}
			A_\nu \\
			a_\nu
		\end{array} \right),
	\end{align}
	where we defined the gauge dynamical operator
	\begin{align}\label{Omunu}
		O^{\mu \nu} = \left( \begin{array}{cc}
			\Box \Theta^{\mu \nu} + 4 e^2 v^2 \eta^{\mu \nu}  & \mu \epsilon^{\mu \rho \nu } \partial_\rho \\
			\mu \epsilon^{\mu \rho \nu } \partial_\rho &   \Box \Theta^{\mu \nu} + 4 g^2 v^2 \eta^{\mu \nu}
		\end{array} \right).
	\end{align}
	After some manipulations, from the inverse of Eq.~\eqref{Omunu}, 
	one can find the dispersion relations $p^2_\pm = m^2_\pm$, where:
	\begin{align}\label{dispersion}
		m^2_\pm &= \frac{1}{2} [\mu^2 + 4 v^2 (e^2+ g^2)] \nonumber \\
		&\pm 
		\frac{1}{2}\sqrt{\left[\mu^2+4v^2(e^2+ g^2)\right]^2-(8v^2eg)^2}.
	\end{align}
	
	It should be stressed that the above relation is necessarily real and non-negative, which ensures the absence of taquions in the model. We can see that the gauge fields will acquire mass contributions coming from the Higgs mechanism and also from the CS term. In particular, in the absence of a CS term ($\mu = 0$), we would have two massive vector bosons with $M_e = 2 e v$ and $M_g = 2 g v$. In the case without spontaneous symmetry breaking ($v=0$), the Higgs mechanism does not take place and we find only a topological mass given by $\mu$. In the absence of a Maxwell term, we obtain two copies of the dispersion relation $p^2 = 16 e^2 g^2 v^4/ \mu^2$, and we have degenerate gauge boson masses.
	
	
	\section{Topological Configurations}\label{sec_topology}

	In order to have finite energy, each non-negative term in Eq.~\eqref{Energy} must asymptote to zero as $\vert \vec{x} \vert = r \rightarrow \infty$. These asymptotic conditions can be seen as boundary conditions for the fields at $S^{1}_\infty \equiv \partial\mathbb{R}^2$ (the circle at infinity). In particular, the scalar fields must asymptote to the vacuum manifold, {\it i.e.}, with a fixed norm on the space of fields, but with phase freedom. In fact, since we have $\phi_+ $ and $\phi_-$, there are two phase degrees of freedom in the asymptotic limit. This give us a map $\Phi_\infty: S^1_\infty \rightarrow S^1 \times S^1 \equiv U(1) \times U(1)$. Any such map can be classified by two integers determined by the fundamental homotopy group $\pi_1\left(S^1 \times S^1\right) \equiv \mathbb{Z} \times \mathbb{Z}$. Therefore we conclude that the finite-energy condition implies an homotopy classification leading to a labeling of the configurations by two integers.
	
	In the asymptotic limit, we can take $\phi_\pm \rightarrow  v e^{i \omega_\pm(\theta)}$ where $\theta$ parametrizes the sphere $S^1_\infty$, together with $A_i \rightarrow - \partial_i\left(\omega_+ + \omega_-\right)/2 e$ and $a_i \rightarrow - \partial_i\left(\omega_+ - \omega_-\right)/2 g$, to ensure that the covariant derivatives vanish at spatial infinity.  To satisfy the remaining asymptotic conditions, we can take $A_0, a_0 \rightarrow 0$ as well as $\partial_i A_0, \partial_ia_0 \rightarrow 0$.
	
	Let us define a (m,n)-vortex as a finite-energy static configuration obeying the boundary conditions stated above with the particular structure:
	\begin{align} 
		\phi_\pm &\rightarrow v e^{i \left(m \pm n\right) \theta}, \nonumber \\
		A_i &\rightarrow -\frac{m}{e} \partial_i \theta, \nonumber \\
		a_i &\rightarrow -\frac{n}{g} \partial_i \theta.
	\end{align} 
	
	Where, in principle, we demand only that $m\pm n \in \mathbb{Z}$, allowing $m$ and $n$ to take simultaneously half-integer values. In the light of the natural doubling of degrees of freedom necessary to ensure parity invariance, the possibility of half-integer numbers should not be worrisome.
	
	From the equations of motion, we already know that there is a relation between charges and magnetic fluxes. But, by definition,	$\Phi = \int \! d^2x \,  \epsilon^{i j} \partial_i A_j = \int_{S^1_\infty} \! dS \, \hat{r}_i \, \epsilon^{i j} A_j.$
	Upon using the asymptotic behavior of the gauge field and the relations $\hat{\theta}_i = \epsilon_{i j} \hat{r}_j $ and $\epsilon_{i j} \epsilon_{ j k} = - \delta_{i k}$, we have, $\Phi =  \int \! d\theta \, r \, \hat{r}_i \, \epsilon^{i j} \left(- \frac{m}{e r} \epsilon_{j k}\hat{r}_k\right) = \frac{2 \pi}{e} m.$
	Analogously for $\chi$.
	
	Thus:
	\begin{align}
		\Phi = \frac{2 \pi}{e} m, \quad \chi = \frac{2 \pi}{g} n.
	\end{align}
	Therefore, we can conclude that besides the magnetic flux associated with one gauge field being proportional to the electric charge of the other, they are all topologically quantized, and can be written as
	\begin{align}\label{charges}
		Q  = \frac{2 \pi}{g} \, \mu \, n,  \quad G = \frac{2 \pi}{e} \, \mu \, m.
	\end{align}
	
	We propose the following (m,n)-vortex {\it ansatz}:	
	\begin{align}\label{Ansatz}
		\phi_\pm &= v \,F_\pm(r) \, e^{i \left(m \pm n\right) \theta}, \nonumber \\
		A_i &= \frac{1}{e r} \left[A(r) - m\right] \hat{\theta}_i,  \nonumber \\
		a_i &= \frac{1}{g r} \left[a(r) - n\right] \hat{\theta}_i,  \nonumber \\
		A_0 &= \frac{1}{e r} \, \alpha(r),  \nonumber \\
		a_0 &= \frac{1}{g r} \, \beta(r).
	\end{align}
	To satisfy the asymptotic conditions, the functions above must satisfy the following boundary conditions:
	\begin{align}\label{boundcond}
		F_\pm(\infty) = 1, \, A(\infty) = a(\infty) = 0.
	\end{align}
	We impose $F_\pm(0) = 0, \, A(0) = m, \, a(0) = n$, and also $ \alpha(0) = \beta(0) = 0$ to avoid a singularity at the origin, except when $m = \pm n$, because in this case one of the scalar profiles can take a non-zero value at the origin.	Under a parity transformation in the vortex configuration, we have $\left(m,n\right) \rightarrow \left(-m,n\right)$, $r \rightarrow r, \theta \rightarrow -\theta - \pi$ and $ F_\pm \rightarrow F_\mp, A \rightarrow -A, a \rightarrow a, \alpha \rightarrow \alpha, \beta \rightarrow -\beta$.  
	
	The energy density functional, considering this {\it ansatz}, can be written as
	\begin{align}\label{key}
		\epsilon &= \! \frac{1}{2 e^2 r^2} \! \left[\dot{A}^2 + \left(\dot{\alpha} - \frac{\alpha}{r}\right)^2 \right] \! + \frac{1}{2 g^2 r^2} \! \left[\dot{a}^2 + \left(\dot{\beta} - \frac{\beta}{r}\right)^2  \right]     \nonumber \\
		&+ \frac{\lambda v^4}{4} \left[(F_+^2 -1)^2 + (F_-^2 -1)^2\right]   \nonumber \\
		&+ \frac{v^2 }{r^2} \left[ F_+^2 \left(\alpha + \beta\right)^2 + F_-^2 \left(\alpha - \beta\right)^2 \right] \nonumber \\
		&  + v^2 \! \left[ \dot{F}_+^2 + \frac{F_+^2}{r^2} \left(A + a\right)^2 + \dot{F}_-^2 + \frac{F_-^2}{r^2} \left(A - a\right)^2\right]   .
	\end{align}
	
	One can also compute the angular momentum of these finite-energy static vortex-like configurations, given by
	\begin{align}
		J = \int \! d^2x \, \epsilon^{ij} r_i T_{0j}.
	\end{align}
	In general, we can write
	\begin{align}
		T_{0 j} &= \epsilon_{j k} \left( E^k B + e^j b\right) \nonumber \\
		&+ 2 \mathbb{Re}\left( D_0 \phi_+^* D_j \phi_+ + D_0 \phi_-^* D_j \phi_-
		\right).
	\end{align}
	Therefore, the angular momentum can be written as a sum of a contribution $J_g$ coming from the gauge fields and another, $J_s$ from the scalar field sector.
	Using the rotationally symmetric {\it ansatz}~\eqref{Ansatz}, we can write
	\begin{align}\label{key}
		\mathbb{Re} \left(D_0\phi_\pm^* D_j \phi_\pm\right) = \frac{  \vert \phi_\pm \vert^2}{r} \left[ e A_0 \pm g a_0 \right] \left[ A \pm a \right] \hat{\theta}_j.
	\end{align}
	But in the static limit we can write for the charge densities, $\rho_\pm = -2 \left(e A_0 \pm g a_0\right) \vert \phi_\pm \vert^2$, and thus,
	\begin{align}
		&2 \mathbb{Re}\left( D_0 \phi_+^* D_j \phi_+ + D_0 \phi_-^* D_j \phi_-
		\right) \nonumber \\
		&= -\frac{1}{r} \left[A \,  \left(\rho_+ + \rho_-\right) + a \,  \left( \rho_+ - \rho_-\right) \right] \hat{\theta}_j.
	\end{align}
	Upon using the Gauss laws~\eqref{GaussLaw}, we obtain for the scalar sector contribution:
	\begin{align}
		J_s = \! \int \! d^2x \! \left[ \frac{A}{e} \left(\nabla \cdot E + \mu b\right) + \frac{a}{g} \left( \nabla \cdot e + \mu B\right)   \right].
	\end{align}
	Now, integrating by parts and using the boundary conditions, this expression will give us a contribution that exactly cancels $J_g$, and another that is entirely given in terms of $A$ and $a$:
	\begin{align}\label{key}
		J_s =  \! &\int \! d^2x \left( B \, r_i E^i + b \, r_i e^i\right) \nonumber \\
		&- \frac{2 \pi \mu}{e g}\left[A(\infty) a(\infty) - A(0) a(0)\right].
	\end{align}
	
	Thus, using the {\it ansatz}, boundary conditions and equations of motion, in the static limit we can obtain for the angular momentum of our (m,n)-vortices:
	\begin{align}\label{key}
		J = \frac{ 2 \pi \mu}{e g} \, n m = \frac{Q G}{ 2 \pi \mu}.
	\end{align}
	We conclude that the angular momentum of these configurations is quantized, proportional to the product of charges, and fractional, exhibiting an anyonic nature. 
	
	Inserting this {\it ansatz} in the equations of motion, we obtain differential equations that must be solved in order to find an explicit solution. 
	From the equations of motion, we obtain:
	\begin{align}
		\ddot{\alpha} - \frac{\dot{\alpha}}{r} + \frac{\alpha}{r^2} + \mu \frac{e}{g} \dot{a}  &= \frac{M_e^2}{2} \left[ \alpha \,  \Delta \! F_+^2 + \beta \,  \Delta \! F_-^2\right],  \\
		\ddot{\beta} - \frac{\dot{\beta}}{r} + \frac{\beta}{r^2} + \mu \frac{g}{e} \dot{A}  &=  \frac{M_g^2}{2} \left[  \beta \,  \Delta \! F_+^2 + \alpha \,  \Delta \! F_-^2\right].  
	\end{align}
	and,
	\begin{align}
		\ddot{A} - \frac{\dot{A}}{r}   + \mu \frac{ e}{g }  \left(\dot{\beta} - \frac{\beta}{r}\right) &= \frac{M_e^2}{2} \left[A \,  \Delta \! F_+^2  + a \, \Delta \! F_-^2 \right],  \\
		\ddot{a} - \frac{\dot{a}}{r}   + \mu \frac{g}{e }  \left(\dot{\alpha} - \frac{\alpha}{r}\right) &= \frac{M_g^2}{2} \left[a \, \Delta \! F_+^2  + A \, \Delta \! F_-^2 \right],
	\end{align}
	where we defined $\Delta \! F_\pm^2 = F_+^2 \pm F_-^2$. The first two equations correspond to the $\nu=0$ components, and the last two to the $\nu=i$ components. From the scalar sector:
	\begin{align}
		\ddot{F}_\pm \!+\! \frac{\dot{F}_\pm}{r} \!+\! \frac{F_\pm}{r^2} \! \left[(\alpha \pm \beta)^2 \! - \! (A \pm a)^2\right] \! = \! \frac{m_S^2}{2}\left(F_\pm^2 - 1\right) \! F_\pm. 
	\end{align}	
	
	These are the differential equations that we need to solve considering the boundary conditions given in Eq.~\eqref{boundcond} and the initial conditions stated in sequence. We were not able to find an analytical solution for these equations, and therefore, in the next section we will present for numerical solutions considering some particular cases that represent different possible scenarios.
	
	In the above differential equations, one can note the appearance of a few mass scales, given by $m_S = \sqrt{\lambda v^2}, \, M_e = 2 e v, \, M_g = 2 g v, \, \text{and finally}, \, \mu$. We can introduce the dimensionless coefficients $K_1 = \mu / m_S, \, K_2 = M_e / M_g = e/g, \, \text{and} \,  K_3 = M_e / m_S$, writing the equations above using the dimensionless distance $x = m_S \, r$ (the derivatives from now on are with respect to $x$), in such a way that the differential equations can be written:  
	
	\begin{align}\label{DiffEqs}
		&\ddot{F}_+ \!+\! \frac{\dot{F}_+}{x} \!+\! \frac{F_+}{x^2}\left[(\alpha + \beta)^2 - (A+a)^2\right] \! = \! \frac{1}{2} \! \left(F_+^2 - 1\right) \! F_+, \nonumber \\
		&\ddot{F}_- \!+\! \frac{\dot{F}_-}{x} \!+\! \frac{F_-}{x^2}\left[(\alpha - \beta)^2 - (A-a)^2\right] \! = \! \frac{1}{2} \! \left(F_-^2 - 1\right) \! F_-, \nonumber \\
		& \ddot{A} - \frac{\dot{A}}{x}  +  K_1 K_2  \left(\dot{\beta} - \frac{\beta}{x}\right) = \frac{K_3^2}{2} \left[A  \, \Delta \! F_+^2  + a \, \Delta \! F_-^2 \right], \nonumber \\
		& \ddot{a} - \frac{\dot{a}}{x}   +  \frac{K_1}{K_2} \left(\dot{\alpha} - \frac{\alpha}{x}\right) = \frac{K_3^2}{2 K_2^2} \left[a \, \Delta \! F_+^2  + A \, \Delta \! F_-^2\right], \nonumber \\
		& \ddot{\alpha} - \frac{\dot{\alpha}}{x} + \frac{\alpha}{x^2} +  K_1 K_2 \dot{a}  = \frac{K_3^2}{2} \left[ \alpha \, \Delta \! F_+^2 + \beta \, \Delta \! F_-^2\right], \nonumber \\
		& \ddot{\beta} - \frac{\dot{\beta}}{x} + \frac{\beta}{x^2} + \frac{K_1}{K_2} \dot{A}  =   \frac{K_3^2}{2 K_2^2} \left[  \beta \, \Delta \! F_+^2 + \alpha \, \Delta \! F_-^2\right].  
	\end{align} 
	
	Before diving headfirst in the numerical solutions for these differential equations, we can briefly analyze the asymptotic behavior of the vortex configurations.
	In fact, considering the asymptotic behaviors for the profiles $F_\pm \rightarrow 1$ and $A,a,\alpha,\beta \rightarrow 0$, we can write $F_\pm = 1 - \tilde{F}_\pm$, $A = 0 + \tilde{A}$, $a = 0 + \tilde{a}$, $\alpha = 0 + \tilde{\alpha}$ and $\beta = 0 + \tilde{\beta}$, where all the quantities with tilde are very small for large $x$. In this regime, we will consider only first order terms in the quantities with tilde, neglecting higher orders.
	
	In this approximation, the first two equations in Eq.~\eqref{DiffEqs} become $ \ddot{\tilde{F}} +  \dot{\tilde{F}}/x - \tilde{F} = 0$, where we already used the expansion described above and neglected higher order terms. Notice that this is a modified Bessel equation, therefore we can write for the asymptotic behavior of the scalar profiles, $F(r) \approx 1 - C \mathcal{K}_0(m_S \, r)$, and 	conclude that the scalar fields will approach their asymptotic value exponentially with a characteristic decay length given by the scalar mass.
	In the same way, we can consider the third and last equations in Eq.~\eqref{DiffEqs}. Using the same approximation discussed above, we obtain the following equations: $\ddot{\tilde{A}} - \frac{\dot{\tilde{A}}}{x} + K_1 K_2 \left(\dot{\tilde{\beta}} - \frac{\tilde{\beta}}{x}\right) = K_3^2 \tilde{A}$ and $\ddot{\tilde{\beta}} - \frac{\dot{\tilde{\beta}}}{x} + \frac{\tilde{\beta}}{x^2} + \frac{K_1}{K_2} \dot{\tilde{A}} = \frac{K_3^2}{K_2^2} \tilde{\beta}$. These differential equations lead to the following asymptotic behavior in terms of the modified Bessel functions of the second kind:
	\begin{align}\label{key}
		A(r) &\approx C_\pm \,r \, \mathcal{K}_1(m_\pm r), \nonumber \\
		\beta(r) &\approx D_\pm \, \mathcal{K}_0(m_\pm \, r).
	\end{align}
	Therefore, the gauge profiles approach their asymptotic value exponentially, with a decay length given by the gauge field masses $m_\pm$, given in Eq.~\eqref{dispersion}. The question of whether both $m_+$ and $m_-$ are equally valid is a subtle one (see Refs.~\cite{JatkarKhare, Inozemtsev, Lozano}), and should be investigated elsewhere. The same analysis can be done with the remaining equations and naturally gives us similar results. 
	
	\section{Explicit vortex solutions}\label{sec_vortex}
	
	In this section we will exhibit explicit numerical solutions for the differential equations presented in the last section. The general strategy adopted here is as follows. We propose to expand the profile functions $F_+, F_-, A, a, \alpha, \beta$ in powers of $x$ around the origin, for example, $A(x) = \sum_{k} A_k x^k$. Plugging these expansions in the above differential equations and using the initial conditions, we can obtain constraints in the expansion coefficients. With these expansions near the origin at hand, we can proceed to search the numerical solutions that will also satisfy the boundary conditions at infinity  using a shooting method. It is important to note that, since we have $A(0)=m, \, a(0)=n$, we need first of all to specify which (m, n)-vortex we are trying to find.
	
	In general lines, for the equations and initial conditions considered here, there are six coefficients to be adjusted; the others vanish or can be found in terms of these six and of the mass quotients $K_i$. Roughly speaking, near the origin  we obtained the following structure of expansions:
	\begin{align}
		F_+(x) &= f_+ \, x^{\vert n + m \vert} + ..., \nonumber \\
		F_-(x) &= f_- \, x^{\vert n - m \vert} + ..., \nonumber \\
		A(x) &= m \! + \! A_2 x^2 \!+ A_+ x^{2 \vert n + m \vert +2} \!+ A_- x^{2 \vert n - m \vert +2}	\!+\! ..., \nonumber \\
		a(x) &= n + a_2 x^2 + a_+ x^{2 \vert n + m \vert +2} + a_- x^{2 \vert n - m \vert +2}	+ \!..., \nonumber \\
		\alpha(x) &= \alpha_1 x + \alpha_+   x^{2 \vert n + m \vert +1} + \alpha_-   x^{2 \vert n - m \vert +1} + ...,  \nonumber \\
		\beta(x) &= \beta_1 x + \beta_+   x^{2 \vert n + m \vert +1} + \beta_-   x^{2 \vert n - m \vert +1} + ...,  
	\end{align}
	where $f_+, f_-, A_2, a_2, \alpha_1, \beta_1$ are free parameters that are determined for each set of $(m, n, K_1, K_2, K_3)$, in order to satisfy the asymptotic conditions at infinity.
	
	In the following, we consider some examples representing distinctive classes of vortices. For each case, we show explicit numerical solutions and analyze some aspects of them, stating the relevant parameters for the solution. In Sec.~\ref{subsecA}, we will analyze the situation where one of the integers is zero, using the case $\left(m=0 , n=1 \right)$ as an example; In Sec.~\ref{subsecB}, we investigate the situation where $m$ and $n$ are equal and non-zero, adopting the case $\left(m = n  = 1\right)$ as illustration, and briefly commenting on $( m= n = 1/2 )$; In Sec.~\ref{subsecC}, we study the case where $m$ and $n$ are non-zero and different, using the case $\left(m=2, n=1\right)$ as an example, and commenting on the case $(m=1/2, n=3/2)$; Finally, in Sec.~\ref{AppendixA}, we analyze solutions obtained with different coefficients $K_i$.

	\subsection{m=0, n=1}\label{subsecA}
	Let us focus first on the solutions with $m=0$ and $n=1$, since this is the simplest possible scenario. In this case, we obtain $\Phi=0$, implying $G=0$ and $J=0$, but $\chi = 2 \pi / g$, giving $Q = 2 \pi \mu / g$. Thus, we would be dealing with configurations without magnetic flux, g-electric charge and angular momentum, but with non-trivial g-magnetic flux and electric charge.
	
\begin{figure}[t!]
		\begin{minipage}[b]{1.0\linewidth}
			\includegraphics[width=\textwidth]{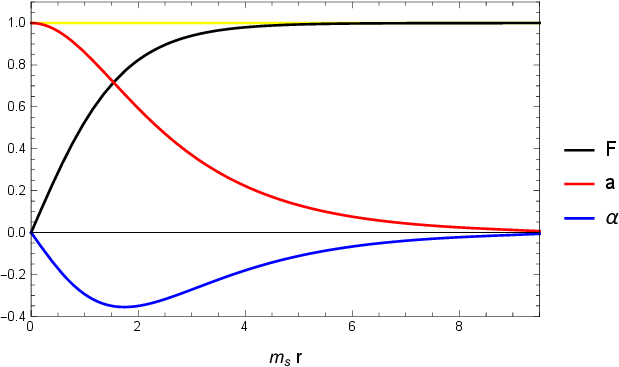}
		\end{minipage} \hfill
		\caption{Vortex solution for $m=0 , n=1$. The scalar profile $F$ is shown in black, and the gauge profiles $a$ and $\alpha$ in red and blue, respectively, as functions of $x = m_S \, r$. The other profiles are identically zero. The relevant parameters here are: $ F_{1}  = 0.58939309 $, $ a_2 = -0.16046967 $,  $ \alpha_1 = -0.36281397 $.}
		\label{solucaon1m0}
	\end{figure}

	Following the procedure described in the beginning of this section, we found a numerical solution for the full set of differential equations that has the property of giving equal profiles $F_+ =  F_-$ and identically zero solutions for $A = \beta = 0$. 
	This means that, for this simple $\left(m=0, n=1\right)$ case, we found {\it a posteriori} that  only half of the differential equations are non-trivial, and therefore  in the numerical analysis we only considered these ones to simplify the analysis. The non-trivial  profiles for the vortex solution are exhibited in Fig.~\ref{solucaon1m0}. 

	\begin{figure}[t!]
		\begin{minipage}[b]{1.0\linewidth}
			\includegraphics[width=\textwidth]{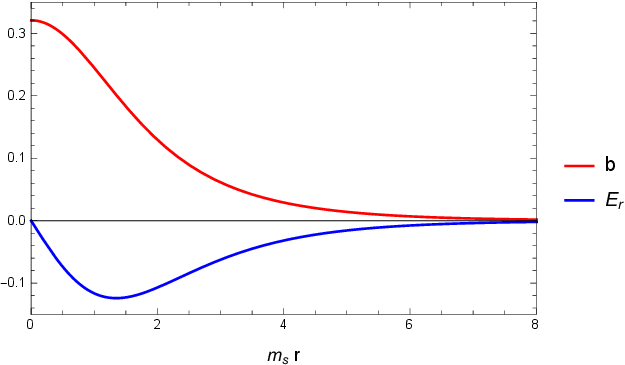}
		\end{minipage} \hfill
		\caption{The g-magnetic (in red) and electric (in blue) fields as functions of $x = m_S \, r$ for the $m=0 , n=1$ solution, in units of $g/m_S^2$ and $e/m_S^2$, respectively.}
		\label{camposEBn1m0}
	\end{figure}
	
	Given this explicit solution, we can immediately plot the g-magnetic and electric fields related with this vortex solution, as one can see in Fig.~\ref{camposEBn1m0}. 
	Notice that the g-magnetic field is finite, non-vanishing, and acquires its maximum value at the origin. The electric field is zero at the origin, maximum at a finite distance and vanishes asymptotically. This is exactly the situation reported in Ref.~\cite{PaulKhare}, where the authors considered an AH model in the presence of a CS term, and obtained a charged vortex solution. This is not a coincidence, because, although physically different, mathematically speaking we are in a similar situation, since we have exactly the same differential equations to be solved.
	But it should be stressed that, besides the parity-invariance of the model and different field content (for instance, we have two gauge fields instead of only one), our vortex solution has zero angular momentum, instead of a non-zero and fractional value as reported in Ref.~\cite{PaulKhare}. 
	The charge and g-current densities display a similar behavior, vanishing at the origin, attaining their maximum value at a finite distance and decaying asymptotically to zero. We remark that an equivalent situation occurs when we consider the case $m=1, n=0$.

	We were not able to find numerical solutions for $m=0$ or $n=0$ with $F_+ \neq F_-$ and $A \neq 0, \beta \neq 0$. It seems that, at least in this simple scenario with vanishing $m$ or $n$, there is a natural trivialization of a sector.
	One might wonder if this trivialization is somehow a consequence of taking the $K_i$ parameters all equal to 1, since they represent quotients between mass scales appearing in our physical system, but it does not seems to be so. In fact, in Sec.~\ref{AppendixA}, we will consider a few numerical solutions for different values of $K_i$,
	and in all cases we obtained similar scalar and gauge profiles, exhibiting the trivialization property reported above.

	\subsection{m=n=1}\label{subsecB}
	
\begin{figure}[t!]
		\begin{minipage}[b]{1.0\linewidth}
			\includegraphics[width=\textwidth]{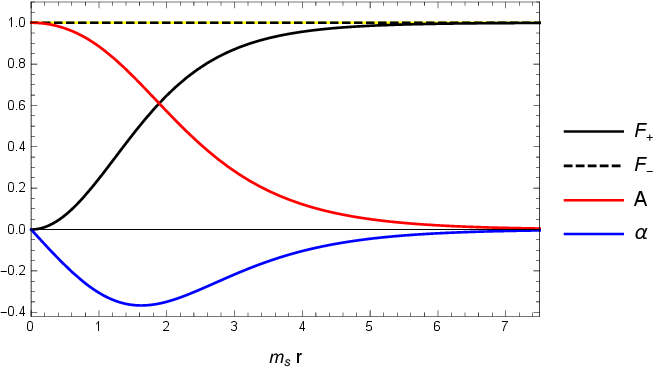}
		\end{minipage}  \hfill
		\caption{Vortex solution for $m = n = 1$. The scalar profile $F_+$ is shown in solid black, $F_-$ in dashed black, and the gauge profiles $a$ and $\alpha$ in red and blue respectively, as functions of $x = m_S \, r$. Notice that here we have $A = a$ and $\alpha = \beta$. The relevant parameters here are $F_{+2} = 0.28684863$, $F_{-0} = 1$, $A_2 = a_2 = -0.10644717$, $\alpha_1 = \beta_1 = -0.36047370$.}
		\label{solucaon1m1}
	\end{figure}

	Now, let us search for solutions with $m = n = 1$. In this case, looking to Eq.~\eqref{charges} we immediately see that $Q = \frac{2 \pi \mu }{g}$ and $G =  \frac{2 \pi \mu }{e}$. This vortex has a non-trivial angular momentum given by $J = \frac{2 \pi \mu}{e g}$, differently from the previous solution. We report this vortex in Fig.~\ref{solucaon1m1}. 
	
	Notice that we obtained {\it a posteriori} a simplified solution where $A = a$, $\alpha = \beta$, and $F_- = 1$. 
	For the scalar profiles, it is important to remember that the exponential part of $\phi_\pm$ involves $m \pm n$. Therefore, the fact that $F_-$ gives us a constant and $F_+$ displays a typical 2-vortex behavior is an indication that the true winding numbers are given by $ m + n $ and $ m - n$, instead of $m $ and $n$ separately.
	
	One can wonder again whether the trivial behavior of the gauge profiles is due to the choice of coefficients. Unlike the previous case, the answer is affirmative, at least with respect to the variation of $K_2$ governing the relationship between different gauge couplings. In fact, starting from the degenerate case and varying $K_2$, the solutions for profiles $A \, \text{and} \, a$ as well as $\alpha \, \text{and} \, \beta$ are not degenerate anymore; however, the scalar profiles do not present any appreciable qualitative change. Varying  $K_1$ and $K_3$, we will find a  behavior similar to the ones described in the last case, as depicted in Sec.~\ref{AppendixA}.

\begin{figure}[t!]
		\begin{minipage}[b]{1.0\linewidth}
			\includegraphics[width=\textwidth]{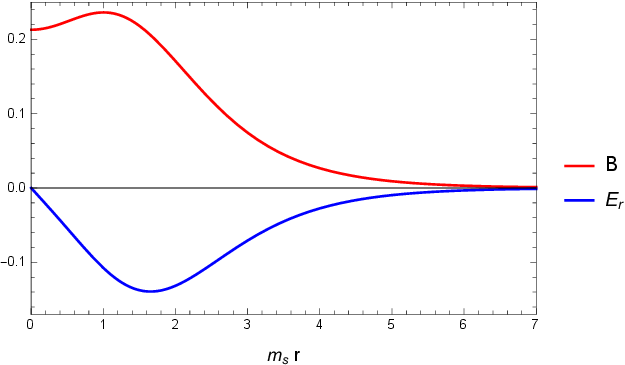}
		\end{minipage}  \hfill
		\caption{The magnetic (in red) and electric (in blue) fields as functions of $x = m_S \, r$ for the $m = n = 1$ solution, in units of $e/m_S^2$. Notice that here we have $B = b$ and $E_r = e_r$.}
		\label{camposEBn1m1}
	\end{figure}

	Given the solution, we can plot its electric and magnetic fields in Fig.~\ref{camposEBn1m1}. 
	The case $m = n  = 1/2$  does not present any appreciable qualitative change in comparison with the solution presented here, except by the scalar profile near the origin, that displays a typical 1-vortex behavior, and by its lowest value of energy and angular momentum $\left( J = \pi \mu / 2 e g  \right)$.
	The energy hierarchy of our solutions will be shortly discussed in the next subsection.
	
	\subsection{m=2, n=1}\label{subsecC}
	
	Finally, we will consider the case $m=2$ and $n=1$. Here, we readily obtain $Q = \frac{2 \pi \mu }{g} $ and $G = \frac{4 \pi \mu }{e} $. Notice that we also have a non-vanishing angular momentum given by $J = \frac{4 \pi \mu }{e g} $. In this case, we expect to see a totally novel result, since there are no simplifications in consequence of the choice of $m$ and $n$. 
	
\begin{figure}[t!]
		\begin{minipage}[b]{1.0\linewidth}
			\includegraphics[width=\textwidth]{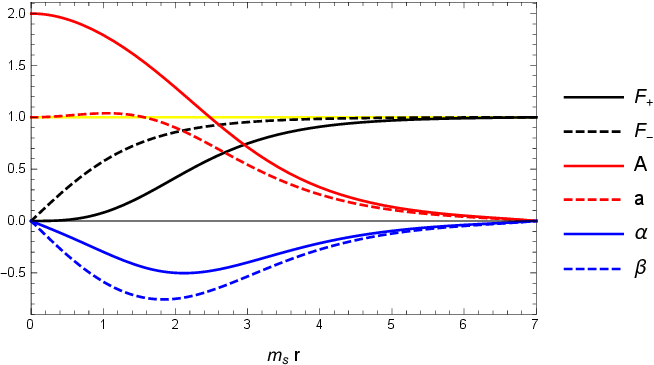}
		\end{minipage}  \hfill
		\caption{Vortex solution for $m=2 , n=1$. The scalar profile $ F_+ $ is shown in solid black, $F_-$ in dashed black; the gauge profile $A$ is shown in solid red, $a$ in dashed red; the profile $ \alpha $ is shown in solid blue, $ \beta $ in dashed blue; all of them are given as functions of $ x = m_S \, r $. The relevant parameters here are $F_{+3} =0.07723697$, $F_{-1} = 0.66377069$, $a_2 = 0.07718614$, $A_2 = -0.22754617$, $\alpha_1 = -0.27824800$, $\beta_1 = -0.68551826$.}
		\label{solucaom2n1}
	\end{figure}

	The numerical solution obtained in this case is given in Fig.~\ref{solucaom2n1}. As one can see, this time there is no degeneracy in the profiles, being all of them non-trivial. In the scalar profiles, notice that $F_-$ displays a behavior near the origin characteristic of a 1-vortex, and $F_+$ of a 3-vortex. 
	
\begin{figure}[t!]
		\begin{minipage}[b]{1.0\linewidth}
			\includegraphics[width=\textwidth]{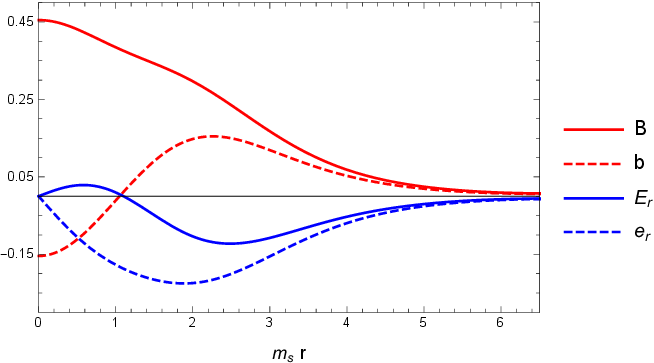}
		\end{minipage}  \hfill
		\caption{The magnetic (solid red) and  electric (solid blue) fields in units of $e/m_S^2$; the  g-magnetic (dashed red) and g-electric (dashed blue) fields in units of $g/m_S^2$. All of them as functions of $x = m_S \, r$ for the $m = 2, n = 1$ solution.}
		\label{camposEBm2n1}
	\end{figure}

	The magnetic and electric fields (as well as the g-magnetic and g-electric) are shown in Fig.~\ref{camposEBm2n1}. 
	For the first time, we observe an oscillating behavior in the electric and g-magnetic fields, and in particular, we see that there is a finite distance where they vanish. 
	Since it is not clear which of the gauge fields (or which combination of them) describes observable electromagnetic phenomena, one should be careful before drawing any conclusion.

	The case $m = 1/2,  n = 3/2$  does not present any appreciable qualitative change in comparison with the solution presented here, except by the scalar profiles near the origin, since $F_+$ and $F_-$ display a behavior typical of 2-vortex and 1-vortex solutions, respectively.

\begin{figure}[t!]
		\begin{minipage}[b]{1.0\linewidth}
			\includegraphics[width=\textwidth]{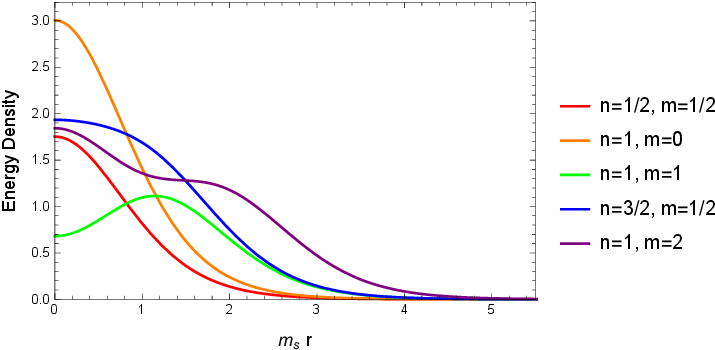}
		\end{minipage}  \hfill
		\caption{The energy density  for the (m,n)-vortex solutions in units of $1/v^2\, m_S^2$. In red, (1/2,1/2); in orange, (0,1); in green, (1,1); in blue, (1/2,3/2); in purple, (2,1). }
		\label{energydensities}
	\end{figure}

	At this point, armed with all these vortex solutions, we can discuss their energy densities and highlight the mass hierarchy between them. Let us first call attention to the fact that we have been successful in finding finite-energy configurations, as one can immediately see in Fig.~\ref{energydensities}. From these energy densities, defining $M_{\left(m,n\right)}$ as the mass associated with the (m,n)-vortex, we obtained the following mass hierarchy in units of $v^2$ : $M_{\left(1/2,1/2\right)} \approx 1.31 < M_{\left(0, 1\right)} \approx 2.27 < M_{\left(1, 1\right)} \approx 2.92 < M_{\left(1/2, 3/2\right)} \approx 3.87 < M_{\left(2, 1\right)} \approx 5.70 $.
	Interestingly enough, one can observe that $M_{\left(1/2,1/2\right)} + M_{\left(-1/2,1/2\right)} = 2 M_{\left(1/2,1/2\right)} > M_{\left(0,1\right)}$. Remember that in the $(\pm 1/2, 1/2)$-vortex, $F_\pm$ is 1-vortex scalar profile, while $F_\mp$ lies in the vacuum, whereas in the $(0, 1)$-vortex both of them are typical 1-vortex scalar profiles. This suggests that there might be an attraction between these vortices. However, to truly understand the interactions between these vortices and conclusively assert this, a more thorough analysis should be done elsewhere, along the lines presented in Ref.~\cite{Jacobs}, for example.
	
	\subsection{Vortex solutions for different $K_i$'s}\label{AppendixA}
	
	In this section, we investigate the existence of vortex solutions and their main properties upon varying the coefficients $K_i$. In the following, we will use as a reference the case $K_1 = K_2 = K_3 = 1$, already studied in the last sections, and change each $K_i$ by a factor of two keeping the others fixed, to find different vortex solutions and compare their main features. 
	
\begin{figure}[t!]
		\begin{minipage}[b]{1.0\linewidth}
			\includegraphics[width=\textwidth]{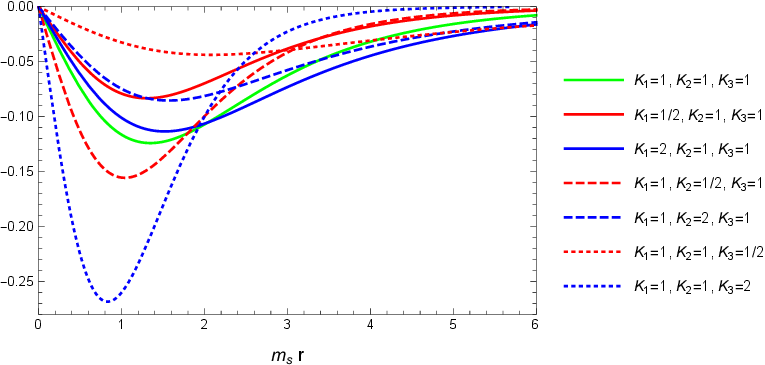}
		\end{minipage} \hfill
		\caption{The electric fields associated with $m=0, n=1$ solution in units of $e/m_S^2$ for different values of $\left( K_1, K_2, K_3 \right)$. In solid green, $\left( 1, 1, 1 \right)$; in solid red, $\left( 1/2, 1, 1 \right)$; in solid blue, $\left( 2, 1, 1 \right)$; in dashed red, $\left( 1, 1/2, 1 \right)$; in dashed blue, $\left( 1, 2, 1 \right)$; in dotted red, $\left( 1, 1, 1/2 \right)$; in dotted blue, $\left( 1, 1, 2 \right)$. }
		\label{Eln1m0variak}
	\end{figure}

\begin{figure}[t!]
		\begin{minipage}[b]{1.0\linewidth}
			\includegraphics[width=\textwidth]{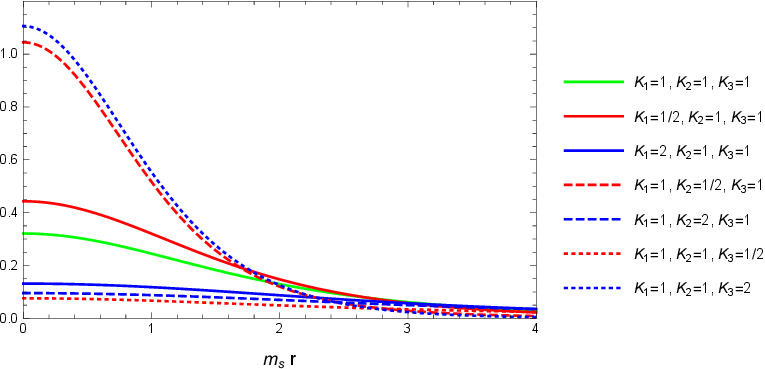}
		\end{minipage}  \hfill
		\caption{The g-magnetic fields associated with $m=0, n=1$ solution in units of $g/m_S^2$ for different values of $\left( K_1, K_2, K_3 \right)$. In solid green, $\left( 1, 1, 1 \right)$; in solid red, $\left( 1/2, 1, 1 \right)$; in solid blue, $\left( 2, 1, 1 \right)$; in dashed red, $\left( 1, 1/2, 1 \right)$; in dashed blue, $\left( 1, 2, 1 \right)$; in dotted red, $\left( 1, 1, 1/2 \right)$; in dotted blue, $\left( 1, 1, 2 \right)$.}
		\label{Magn1m0variak}
	\end{figure}

	Focusing first in the case $m=0, n=1$, the variation of $K_i$ led to qualitatively similar scalar and gauge profiles, and the trivialization property already highlighted before.	
	As one can see from Fig.~\ref{Eln1m0variak}, the electric field qualitative behavior is the same for all the values considered: zero at the origin, attaining a finite non-zero maximum value at some distance and decaying to zero at large distances. Notice that by varying $K_1$, there are only small changes in the profile. By lowering $K_2$, we can observe a more pronounced decay and an improvement in its maximum value. On the other hand, by increasing $K_3$ we observe a sensible increase at the absolute value of the maximum electric field value, accompanied by a more pronounced decay and a small shift in the position where this maximum occur.
	For the g-magnetic field, the qualitative behavior is also the same as we vary $K_i$: attains a finite non-zero maximum value at the origin and decays monotonically as we increase the distance going to zero in the asymptotic limit. By increasing $K_1$, we see that the maximum value of the g-magnetic field diminishes, and this is compatible with the behavior observed in Ref.~\cite{Boyanovsky}. Lowering $K_2$ or increasing $K_3$, we observe a strong change in the maximum value of the g-magnetic field as well as a more pronounced decay as we go far from the origin. Lowering $K_1$, increasing $K_2$, or lowering $K_3$, as before, has the opposite effect, cf. Fig.~\ref{Magn1m0variak}.

\begin{figure}[t!]
		\begin{minipage}[b]{1.0\linewidth}
			\includegraphics[width=\textwidth]{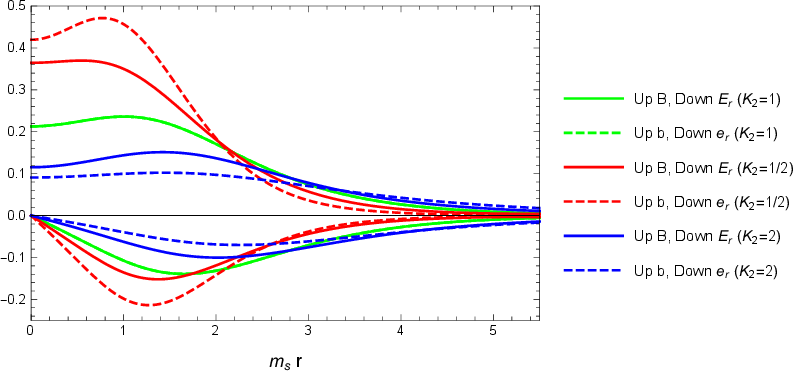}
		\end{minipage}  \hfill
		\caption{The magnetic ($B$), g-magnetic ($b$), electric ($E_r$) and g-electric ($e_r$) fields associated with the $m=n=1$ solutions, for different values of $K_2$. The solid lines refer to $B$ and $E_r$; the dashed lines refer to $b$ and $e_r$. $B$ and $b$ are shown in the upper part; $E_r$ and $e_r$ are shown in the lower part. In green, $K_2 = 1$; in red, $K_2 = 1/2$; in blue, $K_2 = 2$.}
		\label{EBn1m1variak}
	\end{figure}

	Proceeding to the $m = n = 1$ solution, as already highlighted in the main text, the degeneracy that we have found is due to the equality of the couplings when $K_2=1$. When we depart from this simpler case, we find vortex solutions with $A \neq a$ and $\alpha \neq \beta$, naturally leading to different magnetic and g-magnetic (as well as electric and g-electric) fields, as one can see in Fig.~\ref{EBn1m1variak}. Upon varying $K_1$ and $K_3$, we observed the same behavior as described in the previous case. 
	
	Finally, we remark that in the case $m=2 , n=1$ the variation of the coefficients $K_i$ did not lead to any substantial difference from the cases already discussed here. 
	For completeness, it would be interesting to analyze what happens in some limiting cases of this model, for instance, when the CS terms or the Maxwell terms are absent. This analysis is done in the next section.
	
		\section{Vortices in limiting cases}\label{AppendixB}
	
	In this section, we study two particular limits of our model. First, we will briefly address the simpler case in which we do not have a CS term, that is, $\mu = 0$. From a practical point of view, this can be achieved by setting $K_1 = 0$, and the conclusions in this part will come straightforwardly. Notice that this scenario bears resemblance to the usual ANO vortex, since this is nothing but a scalar QED with two gauge fields and two scalars with different charges. 
	
	Second, we will analyze our model in the absence of Maxwell terms, with the gauge kinetic part given solely by the CS term. This allows us to solve the Gauss laws and write the time components of the gauge fields as functions of other quantities. This scenario, where the CS term dominates and the Maxwell terms can be neglected, could be seen as the low-energy regime of our model. 
	
	We remark that the results obtained in this section could be inferred by looking at the behavior of magnetic and electric fields when we changed the coefficient $K_1$ while keeping the others coefficients fixed, since this increases (or decreases) the importance of CS parameter with respect to the other scales of the system. Although it can give us a hint of what would happen in the limits considered here, it is important to remark that the passage from the model considered to the pure CS limit is a subtle one, as one can see for instance in Ref.~\cite{Boyanovsky}, which justifies a separate investigation of the latter.
	
\begin{figure}[t!]
		\begin{minipage}[b]{1.0\linewidth}
			\includegraphics[width=\textwidth]{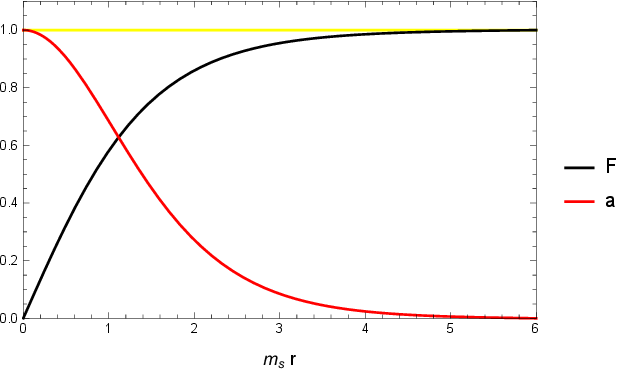}
		\end{minipage}  \hfill
		\caption{Vortex solution for $m = 0, n=1$ in the pure Maxwell limit. The scalar profile $F$ is shown in black and the gauge profile $a$ in red, respectively, as functions of $x = m_S \, r$. The other profiles are identically zero.}
		\label{solucaoMaxwell}
	\end{figure}

\begin{figure}[t!]
		\begin{minipage}[b]{1.0\linewidth}
			\includegraphics[width=\textwidth]{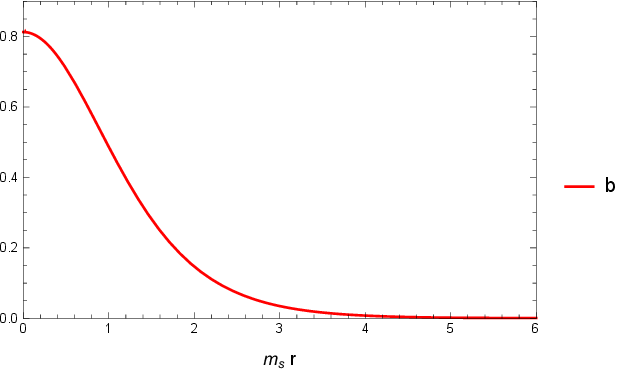}
		\end{minipage}  \hfill
		\caption{The g-magnetic field in the pure Maxwell limit, in units of $g/m_S^2$, as a function of $x = m_S \, r$. The magnetic field as well as the electric and g-electric fields are zero here.}
		\label{EBMaxwell}
	\end{figure}

	Now, we briefly state the results for $K_1 = \mu / m_S = 0$. We will consider the case $m=0$ and $n=1$ with $K_2 = K_3 = 1$ for definiteness, but we would have similar results in the other examples. The vortex solution {\it per se}
	does not exhibit any appreciable change in the profiles $F$ and $a$ as one can see in Fig.~\ref{solucaoMaxwell}. But now we have $\alpha = 0$, and this fact is the most striking difference that appears in this regime. Since we do not have the CS Gauss law constraint anymore, the electric field vanishes and we conclude that the vortex is neutral, as expected. The g-magnetic field in this regime is stronger in magnitude, but exhibit the usual profile, attaining a maximum at the origin and decaying as we increase $x$, as one can see in Fig.~\ref{EBMaxwell}. This is in accordance with the already known results (see for example Ref.~\cite{Boyanovsky}).


	Proceeding to the more interesting scenario in which we can neglect the Maxwell terms, the Gauss laws constraints become much simpler,
	\begin{align}
		\mu b &= e \left(\rho_+ + \rho_-\right), \nonumber \\
		\mu B &= g \left(\rho_+ - \rho_-\right).
	\end{align} 
	Without Maxwell terms, we are able to obtain $A_0$ and $a_0$ directly from the other fields. In fact, we can find:
	\begin{align}
		e A_0 \! &= \! \Lambda \! \left[e B \! \left(\vert \phi_+ \vert^2 - \vert \phi_- \vert^2\right) \! - \! g b \! \left(\vert \phi_+ \vert^2 + \vert \phi_- \vert^2\right)\right] \nonumber  \\
		g a_0 \! &= \! \Lambda \! \left[g b \! \left(\vert \phi_+ \vert^2 - \vert \phi_- \vert^2\right) \! - \! e B \! \left(\vert \phi_+ \vert^2 + \vert \phi_- \vert^2\right)\right],	
	\end{align}
	where we defined $\Lambda \equiv \mu / 8 e g \vert \phi_+ \vert^2 \vert \phi_- \vert^2$ for convenience.
	Plugging the {\it ansatz}, and writing in dimensionless variables using $x = m_S \, r$ and the coefficients $K_i$ as before, we obtain the following expressions for $\alpha$ and $\beta$:
	\begin{align}
		\alpha &=   \frac{K_1 K_2}{2 K_3^2} \frac{1}{F_+^2 F_-^2} \left[\dot{a} \left(F_+^2 + F_-^2\right) -\dot{A} \left(F_+^2 - F_-^2\right)  \right], \nonumber \\
		\beta &=   \frac{K_1 K_2}{2 K_3^2} \frac{1}{F_+^2 F_-^2} \left[ \dot{A} \left(F_+^2 + F_-^2\right) - \dot{a} \left(F_+^2 - F_-^2\right)   \right].
	\end{align}  
	Now, we need only to plug these analytic expressions for $\alpha $ and $\beta$ in the differential equations~\eqref{DiffEqs}, ignoring the contributions coming from the Maxwell terms, and solve them for given $m$ and $n$. Notice that we need only to care about the first four equations, since the last two are already satisfied when we write $\alpha$ and $\beta$ as above.
	
	Although this is a legitimate path to be followed, we simply solved the full set of differential equations in the absence of Maxwell contributions, without using explicitly the CS constraint, stated here only for completeness. In the following, we will exhibit the solution profiles and also the electric and magnetic (as well as g-electric and g-magnetic) fields associated with them. For all of them, we considered $K_1=K_2=K_3=1$ for simplicity.
	
\begin{figure}[t!]
		\begin{minipage}[b]{1.0\linewidth}
			\includegraphics[width=\textwidth]{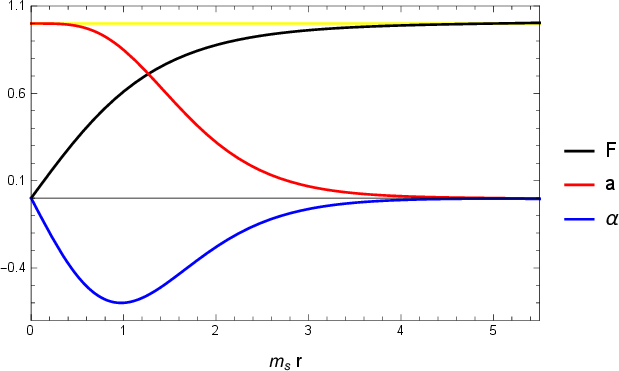}
		\end{minipage}  \hfill
		\caption{Vortex solution for $m = 0, n=1$ in the pure CS limit. The scalar profile $F$ is shown in black; the gauge profiles $a$ and $\alpha$ are shown in red and blue, respectively, as functions of $x = m_S \, r$. The other profiles are identically zero. }
		\label{solucaon1m0CS}
	\end{figure}

\begin{figure}[t!]
		\begin{minipage}[b]{1.0\linewidth}
			\includegraphics[width=\textwidth]{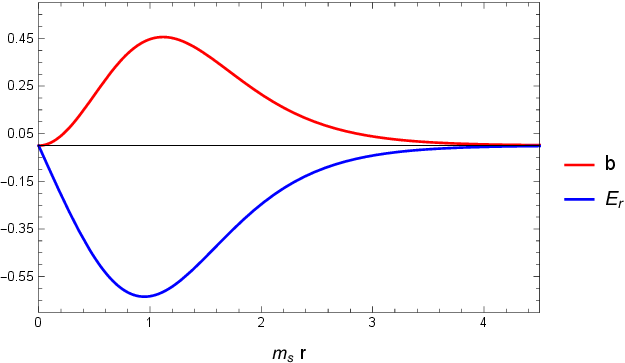}
		\end{minipage}  \hfill
		\caption{The g-magnetic (in red) and electric (in blue) fields as functions of $x = m_S \, r$ for the $m = 0, n = 1$ solution in the pure CS limit, in units of $g/m_S^2$ and $e/m_S^2$, respectively.}
		\label{EBn1m0CS}
	\end{figure}

	The solution for the equations of motion in the pure CS regime for the case $m=0, n=1$ is given in Fig.~\ref{solucaon1m0CS}; the electric and g-magnetic fields are shown in Fig.~\ref{EBn1m0CS}. Notice that they are zero at the origin, attains their maximum value at a finite distance and decays asymptotically, exactly as reported in Ref.~\cite{JackiwWeinberg}, for example. 
	
	
\begin{figure}[t!]
		\begin{minipage}[b]{1.0\linewidth}
			\includegraphics[width=\textwidth]{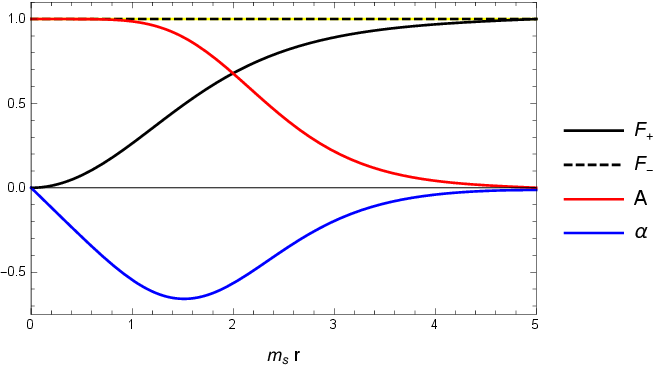}
		\end{minipage}  \hfill
		\caption{Vortex solution for $m = n =1$ in the pure CS limit. The scalar profile $F_+$ is shown in solid black and $F_-$ in dashed black; the gauge profiles $A$ and $\alpha$ are shown in red and blue, respectively, as functions of $x = m_S \, r$. Notice that here we have $A = a$ and $\alpha = \beta$.}
		\label{solucaon1m1CS}
	\end{figure}

\begin{figure}[t!]
		\begin{minipage}[b]{1.0\linewidth}
			\includegraphics[width=\textwidth]{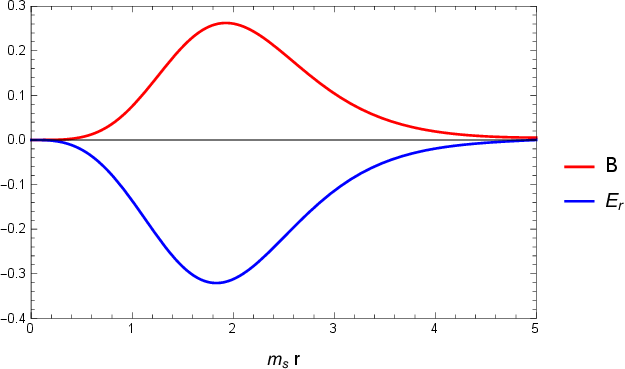}
		\end{minipage}  \hfill
		\caption{The magnetic (in red) and electric (in blue) fields as functions of $x = m_S \, r$ for the $m = n = 1$ solution in the pure CS limit, in units of $e/m_S^2$. Notice that here we have $B=b$ and $E_r = e_r$.}
		\label{EBn1m1CS}
	\end{figure}

	The $m=n=1$ case gives very similar results, see Figs.~\ref{solucaon1m1CS}, \ref{EBn1m1CS}. Remember that we are considering here the particular case in which $K_2 = 1$ and therefore we have degenerate solutions, as we already discussed before.  
	
	
\begin{figure}[t!]
		\begin{minipage}[b]{1.0\linewidth}
			\includegraphics[width=\textwidth]{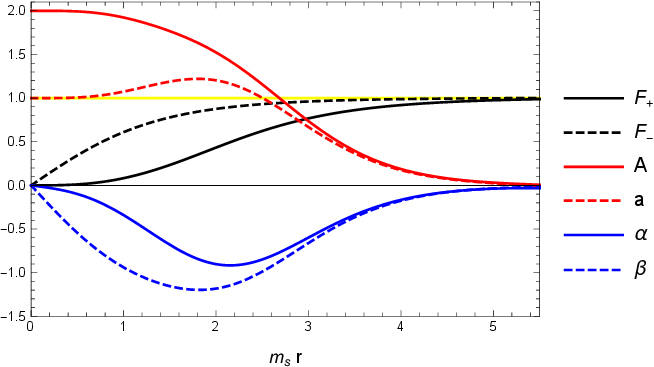}
		\end{minipage}  \hfill
		\caption{Vortex solution for $m=2 , n=1$ in the pure CS limit. The scalar profile $ F_+ $ is shown in solid black, $F_-$ in dashed black; the gauge profile $A$ is shown in solid red, $a$ in dashed red; the profile $ \alpha $ is shown in solid blue, $ \beta $ in dashed blue; all of them are given as functions of $ x = m_S \, r $.}
		\label{solucaon1m2CS}
	\end{figure}

\begin{figure}[t!]
		\begin{minipage}[b]{1.0\linewidth}
			\includegraphics[width=\textwidth]{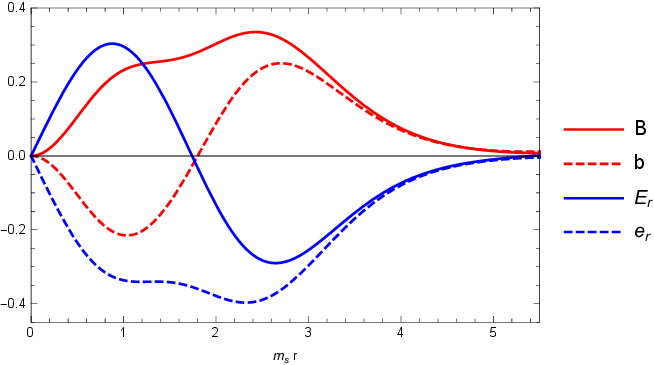}
		\end{minipage}  \hfill
		\caption{The magnetic (solid red) and  electric (solid blue) fields in units of $e/m_S^2$; the  g-magnetic (dashed red) and g-electric (dashed blue) fields in units of $g/m_S^2$. All of them as functions of $x = m_S \, r$ for the $m = 2, n = 1$ solution in the pure CS limit.}
		\label{EBn1m2CS}
	\end{figure}

	The case $m=2, n=1$ presents a more complicated behavior, but it is reminiscent of the solution presented in the main text, as expected. In fact, the solutions are shown in Fig.~\ref{solucaon1m2CS} and the electric and magnetic (as well as g-electric and g-magnetic) fields are shown in Fig.~\ref{EBn1m2CS}. In particular, we still have non-trivial solutions for all profiles and an oscillating behavior for the  fields.	
	

	\section{Conclusions and outlook}\label{sec_conclusions}
	
	In this work, we considered a parity-invariant Maxwell-Chern-Simons $U(1) \times U(1)$ model coupled with charged scalars in 2+1 dimensions, and investigated the existence of topological vortices in this scenario. We described the main features of the model and discussed general properties of topological configurations that could be present in it. Using an appropriate {\it ansatz} and the equations of motion, we obtained the relevant differential equations and solved them numerically. We explicitly analyzed three examples that are representatives of the possible solutions and showed explicit vortex configurations for each case, describing their main properties such as the electric and magnetic fields related with each particular solution.
	We observed the usual expected behavior for vortex configurations, namely: spatially localized configurations with finite energy. Near the origin, the scalar profiles behave as $\phi_\pm (r) \approx r^{\vert n\pm m \vert}$, however when $m \pm n = 0$, $\phi_\pm$ remain constant in their vacuum solution. The electric and g-electric fields both vanish at the origin and asymptote to zero, while in the intermediate region they acquire a maximum absolute value which increases as the ratio $K_2 =\frac{e}{g}$ decreases or $K_3$ increases. The magnetic and g-magnetic fields, despite also asymptoting to zero, only vanish at the origin when $K_1 \propto \mu \rightarrow \infty$, that is to say, in the pure CS limit of the model. The physical properties of the solutions found are summed up on the Table~\ref{TablePhysicProp}.
	
\begin{table}[]
		\renewcommand{\tabcolsep}{4.5 pt}
		\renewcommand{\arraystretch}{1.4}
		\begin{tabular}[t]{|c|c|c|c|}
			\hline
			$(n,m)$ &	$(1,0)$ & $(1,1)$ &  $(1,2)$ \\
			  \hline
			$E(1/v^2)$ & 2.27 & 2.92 & 5.70  \\
			\hline
			$J(eg/2\pi\mu)$ & 0 & 1 & 2 \\
			\hline
			$\Phi (e/2\pi)$ & 0 & 1 & 2    \\
			\hline
			$\chi (g/2\pi)$ & 1 & 1 & 1  \\
			\hline
		\end{tabular}
		\caption{Physical properties of topological vortices for different values of $n$ and $m$.}
		\label{TablePhysicProp}
	\end{table}
	We therefore conclude that there are vortex solutions in this novel class of parity-invariant Maxwell-CS models.

	There are many directions to be explored, for example, it would be interesting to analyze the quantization of the CS parameter, as well as studying this model in a more general manifold. A thorough investigation concerning the interaction between these vortices, answering the question whether they attract or repel, would also be enlightening. The role of monopole operators in this model should be understood, and the possibility of a theory dual to the one presented here could lead to interesting developments.  Furthermore, the product structure of the angular momentum and the presence of two gauge potentials lead us to speculate about a relation between these charged vortices and Dirac monopoles, in the spirit of Refs.~\cite{Cabibbo, Hagen2, Salam}.
	It is of utmost importance  to improve this model allowing the proper investigation of its quantum aspects, to study more general potentials leading to different spontaneous breaking patterns, and also to discuss suitable condensed matter systems that allow this class of models to be experimentally realized.  
	
	An immediate development of this work consists in finding a Bogomoln'yi lower bound for the energy, whose saturation gives rise to first order equations leading to self-dual vortices. In fact, this was done quite recently, see Ref.~\cite{PippoWell}.
	Once this has been accomplished, the investigation of a supersymmetric extension of this model becomes natural, since it is well-known that self-duality and supersymmetry are intimately related~\cite{WittenOlive,DiVecchia,Hlousek1,Hlousek2,LeeLeeWeinberg,LeeLeeMin2,Edelstein,Alvaro}.
	This is work in progress, and the results will be reported soon.
	
	
	\begin{acknowledgments}
		The authors are grateful to J. A. Helay\"el-Neto for helpful discussions, and to P. C. Malta and H. Santos Lima for useful comments.  The authors are also grateful to P. Horvathy, A. Edery and E. Babaev for nice comments and for drawing our attention to some relevant references, and to the anonymous referee for the helpful suggestions.
		The authors also thank the Brazilian scientific support agencies CNPq and FAPERJ for financial support.
		
	\end{acknowledgments}



\begin{thebibliography}{99}
		
			
			\bibitem{Vort1} 
			A. Tonomura {\it et al.}, {\it Motion of vortices in superconductors}, Nature {\bf 397}, 308 (1999).
			
			\bibitem{Vort2} 
			G. Bewley, D. Lathrop, and K. Sreenivasan, {\it Visualization of quantized vortices}, Nature {\bf 441}, 588 (2006).
			
			\bibitem{Vort3} 
			C. Weiler {\it el al.}, {\it Spontaneous vortices in the formation of Bose–Einstein condensates}, Nature {\bf 455}, 948 (2008).
			
			\bibitem{Vort4} 
			S. Autti {\it et al.}, {\it Observation of Half-Quantum Vortices in Topological Superfluid $He^3$}, Phys. Rev. Lett. {\bf 117}, 255301 (2016).
			
			\bibitem{Vort5} 
			G. Gauthier {\it et al.}, {\it Giant vortex clusters in a two-dimensional quantum fluid}, Science {\bf 364}, 1264 (2019).
			
			\bibitem{Vort6} 
			V. Gladilin and M. Wouters, {\it Vortices in Nonequilibrium Photon Condensates}, Phys. Rev. Lett. {\bf 125}, 215301 (2020).
			
			\bibitem{Babaev}	
			E. Babaev, {\it Vortices with fractional flux in two-gap superconductors and in extended Faddeev model}, Phys. Rev. Lett. {\bf 89}, 067001 (2002).
			
			\bibitem{Shifman}
			M. Shifman, {\it Advanced topics in quantum field theory: A lecture course}, Cambridge University Press (2012).
			
			\bibitem{Abrikosov}
			A. A. Abrikosov, {\it On the Magnetic properties of superconductors of the second group}, Sov. Phys. JETP {\bf 5}, 1174 (1957).	
			
			\bibitem{NielsenOlesen}
			H. B. Nielsen and P. Olesen, {\it Vortex-line models for dual strings}, Nucl. Phys. B {\bf 61} 45 (1973).
			
			\bibitem{VegaSchaposnik}
			H. J. de Vega and F. A. Schaposnik, {\it Classical vortex solution of the Abelian Higgs model}, Phys. Rev. D {\bf 14}, 110 (1976).
			
			\bibitem{JuliaZee}
			B. Julia and A. Zee, {\it Poles with both magnetic and electric charges in non-Abelian gauge theory}, Phys. Rev. D {\bf 11}, 2227 (1975).
			
			\bibitem{DeserJackiwTempleton1}
			S. Deser, R. Jackiw, and S. Templeton, {\it Three-Dimensional Massive Gauge Theories}, Phys. Rev. Lett.	{\bf 48},  975 (1982).
			
			\bibitem{DeserJackiwTempleton2}
			S. Deser, R. Jackiw, and S. Templeton, {\it Topologically Massive Gauge Theories},  Ann. Phys. {\bf 140},  372 (1982); Ann. Phys. {\bf 281},  409 (2000).
			
			\bibitem{ChernSimons}
			S.-S. Chern and J. Simons, {\it  Characteristic Forms and Geometric Invariants}, Ann. Math. {\bf 99}, 48 (1974).
			
			\bibitem{Schonfeld}
			J. F. Schonfeld, {\it A mass term for three-dimensional gauge fields}, Nucl. Phys. B {\bf 185}, 157 (1981).
			
			\bibitem{JackiwTempleton}
			R. Jackiw and S. Templeton, {\it How super-renormalizable interactions cure their infrared divergences}, Phys. Rev. D {\bf 23}, 2291 (1981).
			
			\bibitem{HagenAntigo1}
			C. R. Hagen, {\it A New Gauge Theory without an Elementary Photon}, Ann. Phys. {\bf 157}, 342 (1984).
			
			\bibitem{HagenAntigo2}
			C. R. Hagen, {\it What is the most general Abelian gauge theory in two spatial dimensions?}, Phys. Rev. Lett. {\bf 58}, 1074 (1987); Erratum Phys. Rev. Lett. {\bf 58}, 2003 (1987).
			
			\bibitem{Witten}
			E. Witten, {\it Quantum field theory and the Jones polynomial}, Comm. Math. Phys. {\bf 121},  351 (1989).
			
			\bibitem{Dunne}
			G. V. Dunne, {\it Aspects of Chern-Simons theory}, arXiv:9902115.
			
			\bibitem{Horvathy1}
			P. Horváthy and P. Zhang, {\it Vortices in (Abelian) Chern-Simons gauge theory}, Phys. Rept. {\bf 481},  83 (2009).
			
			\bibitem{PaulKhare}
			S. K. Paul and A. Khare, {\it Charged Vortices in Abelian Higgs Model with Chern-Simons Term}, Phys. Lett. B {\bf 174}, 420 (1986).
			
			\bibitem{VegaShap1}
			H. J. de Vega and F. A. Schaposnik, {\it Electrically Charged Vortices in Non-Abelian Gauge Theories with Chern-Simons Term}, Phys. Rev. Lett. {\bf 56}, 2564 (1986). 
			
			\bibitem{VegaShap2}
			H. J. de Vega and F. A. Schaposnik, {\it Vortices and electrically charged vortices in non-Abelian gauge theories}, Phys. Rev. D {\bf 34}, 3206 (1986). 
			
			\bibitem{KumarKhare}
			C. N. Kumar and A. Khare , {\it Charged vortex of finite energy in nonabelian gauge theories with Chern-Simons term}, Phys. Lett. B {\bf 178}, 395 (1986).
			
			\bibitem{Pisarski}
			R. D. Pisarski and S. Rao, {\it Topologically massive chromodynamics in the perturbative regime}, Phys. Rev. D {\bf 32}, 2081 (1985).
			
			\bibitem{Frohlich}
			J. Fr\"{o}hlich and P.A. Marchetti, {\it Quantum field theories of vortices and anyons}, Commun. Math. Phys. {\bf 121}, 177 (1989).
			
			\bibitem{Laughlin}
			R. B. Laughlin, {\it Quantized motion of three two-dimensional electrons in a strong magnetic field}, Phys. Rev. B {\bf 27}, 3383 (1983).
			
			\bibitem{ChenWilczek}
			Y. H. Chen, F. Wilczek, E. Witten, and B. I. Halperin, {\it On anyon superconductivity}, Int. J. Mod. Phys. B {\bf 3}, 1001 (1989).
			
			\bibitem{VolovikYakovenko}
			G. E. Volovik and V. M. Yakovenko, {\it Fractional charge, spin and statistics of solitons in superfluid $He^3$ film},  J. Phys.: Condens. Matter {\bf 1}, 5263 (1989).
			
			\bibitem{JatkarKhare}
			D. P. Jatkar and A. Khare, {\it Peculiar charged vortices in Higgs models with pure Chern-Simons term}, Phys. Lett B {\bf 236}, 283 (1990).
			
			\bibitem{Boyanovsky}
			D. Boyanovsky, {\it Vortices in Landau-Ginzburg theories of anyonic superconductivity}, Nucl. Phys. B {\bf 350}, 906 (1991).
			
			\bibitem{HongKimPac}
			J. Hong, Y. Kim, and P. Y. Pac, {\it Multivortex solutions of the Abelian Chern-Simons-Higgs theory}, Phys. Rev. Lett. {\bf 64}, 2230 (1990).
			
			\bibitem{JackiwWeinberg}	
			R. Jackiw and E. Weinberg, {\it Self-dual Chern-Simons vortices}, Phys. Rev. Lett. {\bf 64}, 2234 (1990).
			
			\bibitem{Bogomol'nyi}
			E. B. Bogomolny, {\it Stability of Classical Solutions }, Sov. J. Nucl. Phys. {\bf 24}, 449 (1976). 
			
			\bibitem{JackiwLeeWeinberg}
			R. Jackiw, K. Lee and E. Weinberg, {\it Self-dual Chern-Simons solitons}, Phys. Rev. D {\bf 42}, 3488 (1990).
			
			\bibitem{WittenOlive}
			E. Witten and D. Olive, {\it Supersymmetry algebras that include topological charges}, Phys. Lett. B {\bf 78}, 97 (1978).
			
			\bibitem{DiVecchia}
			P. Di Vecchia and S. Ferrara, {\it Classical solutions in two-dimensional supersymmetric field theories}, Nucl. Phys. B {\bf 130}, 93 (1977).
			
			\bibitem{Hlousek1}
		Z. Hlousek and D. Spector, {\it Why topological charges imply extended supersymmetry}, Nucl. Phys. B {\bf 370}, 143 (1992).
			
			\bibitem{Hlousek2}
		Z. Hlousek and D. Spector, {\it Bogomol'nyi explained}, Nucl. Phys. B {\bf 397}, 173 (1993).
			
			\bibitem{LeeLeeWeinberg}
			C. Lee, K. Lee, and E. Weinberg,  {\it Supersymmetry and self-dual Chern-Simons systems}, Phys. Lett. B {\bf 243}, 105 (1990).
			
			\bibitem{LeeLeeMin2}
			C. Lee, K. Lee, and H. Min, {\it Supersymmetric Chern-Simons vortex systems and fermion zero modes}, Phys. Rev. D {\bf 45}, 4588 (1992).
			
			\bibitem{Edelstein}
			J. Edelstein, C. Núnez, and F. Schaposnik, {\it Supersymmetry and Bogomol'nyi equations in the Abelian Higgs model}, Phys. Lett. B {\bf 329},  39 (1994).
			
			\bibitem{Alvaro}
			H. R. Christiansen, M. S. Cunha, J. A. Helay\"el-Neto, L. R. U. Manssur, and A. L. M. A. Nogueira, {\it Selfdual vortices in a Maxwell-Chern-Simons model with nonminimal coupling}, Int. J. Mod. Phys. A {\bf 14},  1721 (1999).
			
			\bibitem{LeeLeeMin}
			C. Lee, K. Lee, and H. Min,  {\it Self-dual Maxwell Chern-Simons solitons}, Phys. Lett. B {\bf 252}, 79 (1990).
			
			
			\bibitem{Dunne2}
			G. V. Dunne, {\it Selfdual Chern-Simons theories},  Lect. Notes Phys. M {\bf 36}, 1 (1995).
			
		
			
			\bibitem{JackiwPi1}
			R. Jackiw and S-Y. Pi, {\it Soliton Solutions to the Gauged Nonlinear Schr\"odinger Equation on the Plane}, Phys. Rev. Lett. {\bf 64}, 2969 (1990).
			
			\bibitem{JackiwPi2}
			R. Jackiw and S-Y. Pi, {\it Classical and quantal nonrelativistic Chern-Simons theory}, Phys. Rev. D {\bf 42}, 3500 (1990), (E) {\bf 48}, 3929 (1993).
			
			\bibitem{Manton}
			N. Manton, {\it First Order Vortex Dynamics}, Ann. Phys. {\bf 256}, 114 (1997).
			
			\bibitem{Horvathy2}
			M. Hassa\"ine, P. Horváthy, and J. Yera, {\it Non-relativistic Maxwell-Chern-Simons Vortices}, Ann. Phys. {\bf 263}, 276 (1998).

			
			
			\bibitem{Hagen}
			C. R. Hagen, {\it Parity conservation in Chern-Simons theories and the anyon interpretation}, Phys. Rev. Lett. {\bf 68}, 3821 (1992).
			
			\bibitem{Wilczek}
			F. Wilczek, {\it Disassembling Anyons}, Phys. Rev. Lett. {\bf 69}, 132 (1992).
			
			
			\bibitem{Oswaldo1}
			O. M. Del Cima and E. S. Miranda, {\it Electron-polaron--electron-polaron bound states in mass-gap graphene-like planar quantum electrodynamics: s-wave bipolarons},  Eur. Phys. J. B {\bf 91}, 212 (2018).
			
			\bibitem{Gorbar}
			E. V. Gorbar and S. V. Mashkevich,  {\it Statistical screening in a P-, T-invariant model}, Z. Phys. C - Particles and Fields {\bf 65}, 705 (1995).
			
			\bibitem{Wellisson1}
			W. B. De Lima, O. M. Del Cima, and E. S. Miranda, {\it On the electron–polaron–electron–polaron scattering and Landau levels in pristine graphene-like quantum electrodynamics}, Eur. Phys. J. B {\bf 93}, 187 (2020).
			
			\bibitem{Oswaldo2}
			O. M. Del Cima, D. H. T. Franco, L. S. Lima, and E. S. Miranda, {\it Quantum Parity Conservation in Planar Quantum Electrodynamics}, Int. J. Theor. Phys. {\bf 60}, 3063 (2021).
			
			\bibitem{Wellisson2}
			W. B. De Lima, O. M. Del Cima, and E. S. Miranda, {\it On the ultraviolet finiteness of parity-preserving $\text{U}(1) \times \text{U}(1)$ massive $\text{QED}_3$}, Ann. Phys. {\bf 430},  168504 (2021).
			
			
			\bibitem{Kim}
			C. Kim, C. Lee, P. Ko, B.-H. Lee, and H. Min, {\it Schrodinger fields on the plane with $[U(1)]^N$ Chern-Simons interactions and generalized selfdual solitons}, Phys. Rev. D {\bf 48}, 1821 (1993).
			
			\bibitem{Dziarmaga1}
			J. Diziarmaga, {\it Low-energy dynamics of $[U(1)]^N$ Chern-Simons solitons},  Phys. Rev. D {\bf 49}, 5469 (1994).
		
			\bibitem{Shin2}
			J. Shin, S. Hyun, and J. Yee, {\it Mutual fractional statistics of relativistic Chern-Simons solitons}, Phys. Rev. D {\bf 52},  2591 (1995).
			
			\bibitem{Dziarmaga2}
			J. Diziarmaga, {\it Only hybrid anyons can exist in broken symmetry phase of nonrelativistic $[U(1)]^2$ Chern-Simons theory}, Phys. Rev. D {\bf 50}, R2376(R) (1994).
			
			\bibitem{Shin1}
			J. Shin and J. Yee, {\it Vortex solutions of parity invariant Chern-Simons gauge theory coupled to fermions},  Phys. Rev. D {\bf 50},  4223 (1994).
			
			\bibitem{Exp1}
			R. F. Keifl {\it et al.}, {\it Search for anomalous internal magnetic fields in high-$T_c$ superconductors as evidence for broken time-reversal symmetry}, Phys. Rev. Lett. {\bf 64}, 2082 (1990).
			
			\bibitem{Exp2}
			S. Spielman {\it et al.}, {\it Test for nonreciprocal circular birefringence in $YBa_2Cu_3O_7$ thin films as evidence for broken time-reversal symmetry}, Phys. Rev. Lett. {\bf 65}, 123 (1990).
			
			
			\bibitem{Exp3}
			K. Lyons {\it et al.}, {\it Search for circular dichroism in high-$T_c$	superconductors}, Phys. Rev. Lett. {\bf 64}, 2949 (1990).
			
			\bibitem{Semenoff}
			G. W. Semenoff and N. Weiss, {\it 3D field theory model of a parity invariant anyonic superconductor}, Phys. Lett. B {\bf 250}, 117 (1990). 
			
			\bibitem{Mavromatos}
			N. Dorey and N. E. Mavromatos, {\it Superconductivity in 2+1 dimensions without parity or time-reversal violation}, Phys. Lett. B {\bf 250},  107 (1990). 
			
			
			\bibitem{Kovner}
			A. Kovner and B. Rosenstein, {\it Kosterlitz-Thouless mechanism of two-dimensional superconductivity}, Phys. Rev. B {\bf 42}, 4748 (1990).
			
			\bibitem{Mavromatos2}
			N. Dorey and N. E. Mavromatos, {\it QED3 and two-dimensional superconductivity without parity violation}, Nucl. Phys. B {\bf 386}, 614 (1992).
			
			
		
			\bibitem{Math1}
			C.-S. Lin and J. Prajapat, {\it Vortex Condensates for Relativistic Abelian Chern-Simons Model with Two Higgs Scalar Fields and Two Gauge Fields on a Torus}, Commun. Math. Phys. {\bf 288}, 311 (2009).
			
		
			\bibitem{Math3}
			H.-Y. Huang, Y. Lee, and C.-S. Lin, {\it Uniqueness of topological multi-vortex solutions for a skew-symmetric Chern-Simons system}, J. Math. Phys. {\bf 56}, 041501 (2015).
			
			\bibitem{Math4}
			B. Guo and F. Li, {\it Doubly periodic vortices for a Chern–Simons model}, J. Math. Anal. Appl. {\bf 458}, 889 (2018).
			
			
			
				\bibitem{Kou1}
			S.-P. Kou, X.-L. Qi, and Z.-Y. Weng, {\it Mutual Chern-Simons effective theory of doped antiferromagnets}, Phys. Rev. B {\bf 71}, 235102 (2005).
			
			\bibitem{Kou2}
			S.-P. Kou, M. Levin, and X.-G. Wen, {\it Mutual Chern-Simons theory for $Z_2$ topological order}, Phys. Rev. B {\bf 78}, 155134 (2008).
			
			\bibitem{Kou3}
			S.-P. Kou, X.-L. Qi, and Z.-Y. Weng, {\it Spin Hall effect in a doped Mott insulator}, Phys. Rev. B {\bf 72}, 165114 (2005).
			
			\bibitem{Kou4}
			S.-P. Kou, J. Yu, and X.-G. Wen, {\it Mutual Chern-Simons Landau-Ginzburg theory for continuous quantum phase transition of $Z_2$ topological order}, Phys. Rev. B {\bf 80}, 125101 (2009).
			
			\bibitem{Qi}
			X.-L. Qi and Z.-Y. Weng, {\it Mutual Chern-Simons gauge theory of spontaneous vortex phase}, Phys. Rev. B {\bf 76}, 104502 (2007).
			
			\bibitem{Ye}
			P. Ye, L. Zhang, and Z.-Y. Weng, {\it Superconductivity in mutual Chern-Simons gauge theory}, Phys. Rev. B {\bf 85}, 205142 (2012).
			
			
			\bibitem{Diamantini1}
			M. C. Diamantini, P. Sodano, and C. A. Trugenberger, {\it Self-duality and oblique confinement in planar gauge theories}, Nucl. Phys. B {\bf 448},  505 (1995).
			
			\bibitem{Diamantini2}
			M. C. Diamantini, P. Sodano, and C. A. Trugenberger, {\it Gauge theories of Josephson junction arrays}, Nucl. Phys. B {\bf 474},  641 (1996).
			
			\bibitem{Diamantini3}
			M. C. Diamantini, P. Sodano, and C. A. Trugenberger, {\it Superconductors with topological order}, Eur. Phys. J. B {\bf 53}, 19 (2006).
			
			\bibitem{Sakhi}
			S. Sakhi, {\it Tricritical behavior in the Chern-Simons-Ginzburg-Landau theory of self-dual Josephson junction arrays}, Phys. Rev. D {\bf 97}, 096015 (2018). 
			
			
			
			
			
			
			
			\bibitem{Shaposhnikov}
			M. M. Anber, Y. Burnier, E. Sabancilar, and M. Shaposhnikov, {\it Confined vortices in topologically massive $U(1) \times U(1)$ theory}, Phys. Rev. D {\bf 92}, 065013 (2015).
			
			\bibitem{ColemanHill}
			S. R. Coleman and B. R. Hill, {\it No more corrections to the topological mass term in $QED_3$}, Phys. Lett. B {\bf 159}, 184 (1985).	
			
			
			\bibitem{Penin1}
			A.A. Penin and Q. Weller, {\it What Becomes of Giant Vortices in the Abelian Higgs Model}, Phys. Rev. Lett. {\bf 125} 251601 (2020). 
			
			\bibitem{Penin2}
			A. A. Penin and Q. Weller,  {\it A theory of giant vortices}, J. High Energ. Phys. {\bf 2021}, 56 (2021).
			
			\bibitem{GiantExp1}
			P. L. Marston and W. M. Fairbank, {\it Evidence of a Large Superfluid Vortex in $\text{He}^4$}, Phys. Rev. Lett. {\bf 39}, 1208 (1977).
			
			\bibitem{GiantExp2}
			P. Engels {\it et al.}, {\it Observation of Long-Lived Vortex Aggregates in Rapidly Rotating Bose-Einstein Condensates}, Phys. Rev. Lett. {\bf 90}, 170405 (2003).
			
			\bibitem{GiantExp3}
			T. Cren {\it et al.}, {\it Vortex Fusion and Giant Vortex States in Confined Superconducting Condensates}, Phys. Rev. Lett. {\bf 107}, 097202 (2011).
			
			\bibitem{Helayel}
			Cristine N. Ferreira, J. A.  Helay\"el-Neto, Álvaro L. M. A.  Nogueira, and A. A. V. Paredes, {\it Vortex Formation in a $U(1) \times U(1)'-\mathcal{N}=2-D=3$ Supersymmetric Gauge Model}, PoS {\bf ICMP2013},  011 (2013).  
			
			\bibitem{Edery}
			A. Edery, {\it Non-singular vortices with positive mass in 2+1-dimensional Einstein gravity with $AdS_3$ and Minkowski background}, J. High Energ. Phys. {\bf 2021}, 166 (2021).
			
			\bibitem{Albert}
			J. Albert, { \it The Abrikosov vortex in curved space}, J. High Energ. Phys. {\bf 2021}, 12 (2021).
			
			\bibitem{Schaposnik1}
			P. Arias, A. Arza, F. A. Schaposnik, D. Vargas-Arancibia, and M. Venegas, {\it Vortex solutions in the presence of Dark Portals}, Int. J. Mod. Phys. A {\bf 37}, 2250087 (2022).
			
			\bibitem{Schaposnik2}
			A. Rapoport and F. A. Schaposnik, {\it A d=3 dimensional model with two U(1) gauge fields coupled via matter fields and BF interaction}, Phys. Lett. B {\bf 806}, 135472 (2020).
			
			\bibitem{Schaposnik3}
			G. S. Lozano and F. A. Schaposnik, {\it Vortices in fracton type gauge theories}, Phys. Lett. B {\bf 811}, 135978 (2020).
			
			
			\bibitem{Bazeia1}
			D. Bazeia, M. A. Liao, and M. A. Marques, {\it Generalized Maxwell-Higgs vortices in models with enhanced symmetry}, arXiv: 2201.12115.		
			
			\bibitem{Bazeia2}
			I. Andrade, D. Bazeia, M.A. Marques, and R. Menezes, {\it Long range vortex configurations in generalized models with the Maxwell or Chern-Simons dynamics}, Phys. Rev. D {\bf 102}, 025017 (2020). 
			
			\bibitem{Bazeia3}
			D. Bazeia, M. A. Liao, M. A. Marques, and R. Menezes, {\it Multilayered vortices}, Phys. Rev. Research {\bf 1}, 033053 (2019).
			
			
		
			\bibitem{Nemeth}
			Z. Németh, {\it Remarks on the solutions of the Maxwell-Chern-Simons theories}, Phys. Rev. D {\bf 58}, 067703 (1998).	
			
			
			\bibitem{Binegar}
			B. Binegar, {\it Relativistic field theories in three dimensions}, J. Math. Phys. {\bf 23}, 1511 (1982).
			
			
			\bibitem{Inozemtsev}
			V. I. Inozemtsev, {\it On Charged Vortices in the (2 + 1)-Dimensional Abelian Higgs Model}, EPL {\bf 5}, 113 (1988).
			
			\bibitem{Lozano}
			G. Lozano, M. V. Manias, and F. A. Schaposnik, {\it Charged-vortex solution to spontaneously broken gauge theories with Chern-Simons term}, Phys. Rev. D {\bf 38},  601 (1988).
			
			\bibitem{Jacobs}
			L. Jacobs, A. Khare, C. N. Kumar, and S. K. Paul, {\it The interaction of Chern-Simons vortices}, Int. J. Mod. Phys. A {\bf 6}, 3441 (1991).
		
			
			\bibitem{Cabibbo} 
			N. Cabibbo and E. Ferrari, {\it Quantum electrodynamics with Dirac monopoles}, Nuovo Cim. {\bf 23},  1147 (1962).
			
			\bibitem{Hagen2} 
			C. R. Hagen, {\it Noncovariance of the Dirac Monopole}, Phys. Rev.   {\bf 140}, B804 (1965).
			
			\bibitem{Salam} 
			A. Salam, {\it Magnetic monopole and two photon theories of C violation}, Phys. Lett. {\bf 23}, 683 (1966).
			
			\bibitem{PippoWell}
			W. B. De Lima and P. De Fabritiis, {\it Self-dual Maxwell-Chern-Simons solitons in a parity-invariant scenario}, Phys. Lett. B {\bf 833}, 137326 (2022). 
			
		
			

		
			
			
		\end{thebibliography}
	\end{document}